\documentclass[preprint2]{aastex}

\usepackage{booktabs}
\usepackage{soul}
\usepackage{color}

\begin{document}

\title{Clustering Properties of Far-Infrared Sources in Hi-GAL \\Science Demonstration Phase Fields}

\shorttitle{Clustering properties of Hi-GAL SDP sources.}
\shortauthors{Billot et al.}


\author{N. Billot\altaffilmark{1}, E. Schisano\altaffilmark{2}, M. Pestalozzi\altaffilmark{2}, S. Molinari\altaffilmark{2}, A. Noriega-Crespo\altaffilmark{3}, J.~C.~Mottram\altaffilmark{4}, L.~D.~Anderson\altaffilmark{5}, D. Elia\altaffilmark{2}, G. Stringfellow\altaffilmark{6}, M.~A. Thompson\altaffilmark{7}, D. Polychroni\altaffilmark{2}, L. Testi\altaffilmark{8}}

\altaffiltext{1}{NASA Herschel Science Center, California Institute of Technology, 770 S. Wilson Ave, Pasadena, CA 91125, USA\\ \emph{nbillot@ipac.caltech.edu}}
\altaffiltext{2}{INAF-IFSI, Roma, Italy}
\altaffiltext{3}{Spitzer Science Center, Caltech, Pasadena, USA}
\altaffiltext{4}{School of Physics, University of Exeter, Exeter, Devon, EX4 4QL, UK}
\altaffiltext{5}{Laboratoire d'Astrophysique de Marseille (UMR 6110), Marseille, France}
\altaffiltext{6}{Center for Astrophysics \& Space Astronomy, University of Colorado, USA}
\altaffiltext{7}{Centre for Astrophysics Research, Science \& Technology Research Institute, University of Hertfordshire, College Lane, Hatfield, AL10 9AB, UK}
\altaffiltext{8}{European Southern Observatory, Garching, Gemany}

\begin{abstract}

We use a Minimum Spanning Tree algorithm to characterize the spatial distribution of Galactic Far-IR sources and derive their clustering properties. We aim to reveal the spatial imprint of different types of star forming processes, e.g. isolated spontaneous fragmentation of dense molecular clouds, or events of triggered star formation around \textsc{Hii} regions, and highlight global properties of star formation in the Galaxy. 
We plan to exploit the entire Hi-GAL survey of the inner Galactic plane to gather significant statistics on the clustering properties of star forming regions, and to look for possible correlations with source properties such as mass, temperature or evolutionary stage. In this paper we present a pilot study based on the two 2\degr$\times$2\degr~fields centered at longitudes $l=30$\degr~and $l=59$\degr~obtained during the Science Demonstration Phase (SDP) of the Herschel mission. We find that over half of the clustered sources are associated with \textsc{Hii} regions and infrared dark clouds. Our analysis also reveals a smooth chromatic evolution of the spatial distribution where sources detected at short-wavelengths, likely proto-stars surrounded by warm circumstellar material emitting in the far-infrared, tend to be clustered in dense and compact groups around \textsc{Hii} regions while sources detected at long-wavelengths, presumably cold and dusty density enhancements of the ISM emitting in the sub-millimeter, are distributed in larger and looser groups.


\end{abstract}


\section{Introduction}

Observers usually rely on fitting Spectral Energy Distributions (SEDs), or other distinctive spectral features, with theoretical models to probe the physical properties of Young Stellar Objects (YSOs), and possibly learn about star formation processes \citep[e.g., ][]{shu, genzel, andre93, andre00, evans}. Likewise, the spatial distribution of YSOs contains valuable information about star formation, in particular the imprint of gravitational fragmentation in molecular clouds \citep{gomez, hartmann, allen, schmeja}, or the spatial segregation between sources of different mass \citep{kirk} or evolutionary stage \citep{gutermuth, carlson}. The analysis of clustering properties is in fact complementary to the spectral approach in the sense that it relates to an entire population of objects rather than individual sources. This however calls for a very large sample of YSOs. In addition, to probe the initial spatial distribution of forming stars, or rather star clusters \citep{lada}, it is best to observe the youngest population of YSOs before they diffuse away from their stellar nursery while the gas evaporates from the disrupting cloud \citep{bastian, proszkow}.

We plan to use the \emph{Herschel Space Observatory} \citep{pilbratt}, operating in the far-Infrared/sub-millimeter regime and covering the peak emission of the youngest YSOs, to study their clustering properties and tentatively relate the observed spatial distribution with different mechanisms of star formation. We will also search for correlations between clustering and YSO physical properties - such as mass, temperature, evolutionary stage - as well as YSOs immediate environment - density, radiation field, \textsc{Hii} region. 

Herschel will observe over 270 square degrees of the inner Galactic plane as part of the Hi-GAL survey \citep{molinari}. With its unprecedented angular resolution, sensitivity and spatial coverage, we expect Hi-GAL to detect tens of thousands of sources. This will constitute a very large data set that should enable us to reach high statistical significance for identifying YSO clustering trends. In this article we present a pilot study of our project based on the two 2\degr$\times$2\degr~fields centered in the Galactic plane at longitudes $l=30$\degr~and $l=59$\degr~obtained during the Science Demonstration Phase (SDP) of the Herschel mission. The scope of this article is therefore more modest than if using the entire survey due to the lower statistics available (only 3\% of the survey was covered during the Herschel SDP), yet it presents the methodology and associated diagnostic tools we have developed in preparation to exploiting the entire survey.

In section~\ref{sec:obs}, we present the observations of the two SDP fields and the catalog of extracted sources we use in our analysis. In section~\ref{sec:mst}, we describe our approach to characterize the spatial distribution of Hi-GAL sources using a Minimum Spanning Tree algorithm and exploiting the heliocentric distance information recently obtained by the Hi-GAL consortium \citep{russeil} We discuss our results in section~\ref{sec:results}, in particular the wavelength dependence we find on the clustering properties of the YSOs, as well as the relation between clustering and cloud fragmentation, or the presence of \textsc{Hii} regions. Finally we give our conclusions and prospects for the entire Hi-GAL data set in section~\ref{sec:concl}.

\section{Observations and Source Catalog}
\label{sec:obs}

The \emph{Herschel Space Observatory} \citep{pilbratt} observed two $2.1 \times 2.1$ square degree fields as part of the Science Demonstration Phase of the mission in November~2009. These two target fields were chosen from the Herschel Infrared GALactic plane survey \citep[Hi-GAL, ][]{molinari}, and are approximately located on the Galactic plane at longitudes $l=30$\degr~and $l=59$\degr. 
Observations were carried out in the SPIRE/PACS parallel mode \citep{griffin, poglitsch} at fast scan speed (60\arcsec/s) in two orthogonal directions. This observing strategy provides simultaneous imaging in five bands centered at 70, 160, 250, 350, and 500~$\mu$m, and an angular resolution varying from $\sim$10\arcsec~to 40\arcsec. The cross-scanned observations are used to preserve the extended emission from the interstellar medium (ISM) during the map-making process. Maps are created with the ROMAGAL algorithm (Traficante et al., submitted), and three-color images are presented in figure~1 and~2 of \citet{molinari}. 

The morphology of the $l=30$\degr~field in Herschel bands is mostly shaped by two luminous massive star-forming complexes, namely the mini-starburst W43 \citep{motte, bally} located in the inner arm of our Galaxy at $\sim$5.8~kpc from the Sun, and the ultra compact \textsc{Hii} region G29.96-0.02 \citep[hereafter G29, ][]{beuther} located at $\sim$8~kpc. \citet{paladini} find an additional 23~\textsc{Hii} regions in this 4~square degree field. The $l=59$\degr~field is dominated by the large OB~association Vul~$\!$OB1 \citep{billot}, located in the Sagittarius arm at a distance of 2.3~kpc.

Source detection at these wavelengths is a complex task due to the rich structured backgrounds present in the Galactic plane. \citet{molinari10b} have developed a method, based on the second derivatives of the maps, that filters out low spatial frequencies along multiple directions and reveals compact sources that exhibit strong signal gradients in the image. The photometry is then measured by fitting multiple Gaussians to the detected groups of pixels that have a second derivative value above a given threshold (see \citeauthor{molinari10b} for details). The source extraction is carried out independently at the five wavelengths. Sources detected at 70, 160, 250, 350 and 500~$\mu$m are then band-merged, following the method described in \citet{elia}, to form the source catalog that we use as a starting point for the spatial distribution analysis presented in this article. Note that we will use the term \emph{single-wavelength source catalogs} in section~\ref{sec:results} when referring to single columns of the complete band-merged catalog, i.e. all the sources detected from a single image at a particular wavelength. Table~\ref{tab:entries} gives the numbers of sources extracted in both fields and in each Herschel band individually, as well as the total number of sources in the band-merged catalog.

Furthermore, there has been a significant effort within the Hi-GAL consortium to measure the distance to most sources detected in the SDP fields in order to derive physical parameters such as the mass or the luminosity from the measured fluxes. Distances were estimated using a multi-wavelength approach, exploiting both spectral line emission (kinematic distances) as well as extinction maps, parallax measurements, and physical connection with objects at known distances. Details of the methodology are described in \citet{russeil}. In total, over 2000 sources ($>$90\%) possesses a distance estimate.

\begin{deluxetable}{ccccccc}\centering
	\tablecolumns{8}
	\tablewidth{0pt}
	\tablecaption{Number of entries in the source catalog per band and per field.\label{tab:entries}}
	\tablehead{\colhead{Band} & \colhead{70~$\mu$m} & \colhead{160~$\mu$m} & \colhead{250~$\mu$m} & \colhead{350~$\mu$m} & \colhead{500~$\mu$m} & \colhead{Total\tablenotemark{a}}}
	\startdata
	$l=30$\degr & 698 & 679 & 758 & 785 & 592 & 1565 \emph{(1388)} \\
	$l=59$\degr & 336 & 389 & 675 & 578 & 515 & 1113 \emph{(718)}
	\enddata
\tablenotetext{a}{Indicates the total number of entries in the band-merged catalog having at least one detection in one of the Herschel bands. The slanted number in parenthesis indicates the number of sources for which a distance estimate is available.}
\end{deluxetable}

\section{Characterization of the Spatial Distribution}
\label{sec:mst}

Several mathematical tools are available to characterize the spatial distribution of a set of localized points, e.g. the two-point correlation function, the nearest-neighbor filtering, the minimum spanning tree, the mean surface density of companion, the Voronoi tessellation, etc. All these methods have been successfully applied to astronomical data sets to detect YSO clusters in star forming regions \citep{gomez, hartmann, karr, cartwright, schmeja06, chavarria, gutermuth}. The process of cluster identification always requires the determination of a threshold to isolate source overdensities from the underlying population of distributed objects. This threshold can take the form of a cutoff source surface density or a cutoff source separation, depending on the chosen approach, and it plays a crucial role in deriving cluster properties. Here the term \emph{cluster} is used to designate source overdensities that contrast against the distribution of field objects. In the following analysis, the assignment of cluster membership is solely based on morphological grounds, without any kinematic information. Consequently, the detected clusters might be gravitationally bound entities as well as loose associations simply tracing regions of star formation as defined by \citet{gieles}.
 
In general the characterization of spatial distributions remains vague and imprecise. For instance, from a census of recent studies, \citet{bressert} point out that the fraction of sources in clusters can vary from 40~to 90\% depending on the adopted definition of a cluster. Furthermore, \citet{schmeja10} has conducted a quantitative comparison of 4~different algorithms to identify star clusters in a field, and they all exhibit variable efficiencies depending on the size and character of the investigated area and the purpose of the study.

In the present article, we follow the methodology presented in \citet{gutermuth} to study the spatial distribution of the Hi-GAL data set. \citeauthor{gutermuth} argue in favor of the Minimum Spanning Tree (MST) algorithm for several reasons: (1) the `overdensity' threshold is derived in a systematic manner from the data itself, (2) it creates fully connected entities rather than islands of isolated groups with few sources as with the nearest-neighbor approach, and (3) there is no inherent smoothing associated with the MST analysis so that there is no bias with regard to the shapes of the clusters one can isolate. 

\subsection{The Control Distribution}
\label{subsec:control}

\begin{figure*}
  \centering
    \begin{tabular}{cc}
      \includegraphics[width=0.45\textwidth]{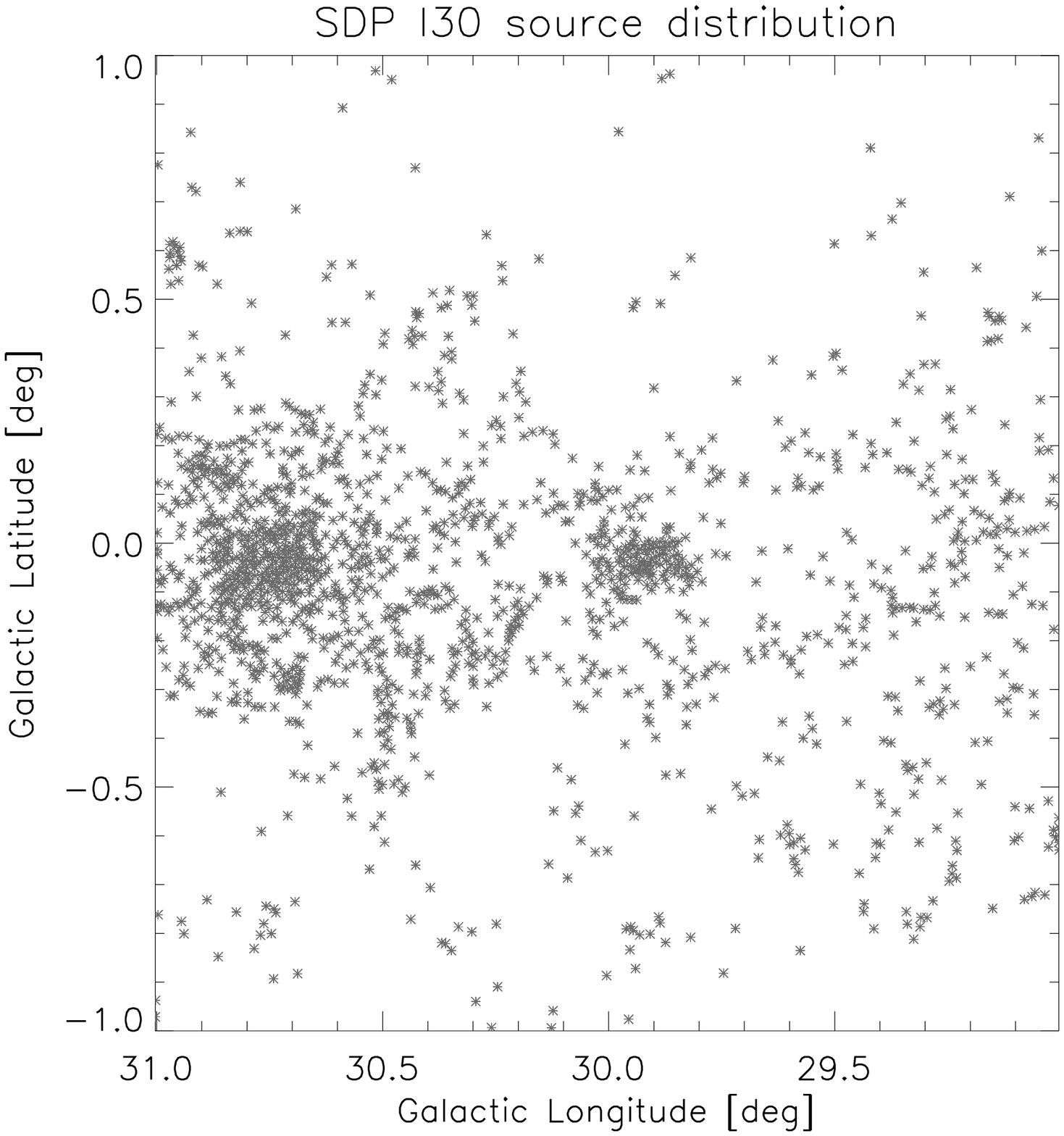}&
      \includegraphics[width=0.45\textwidth]{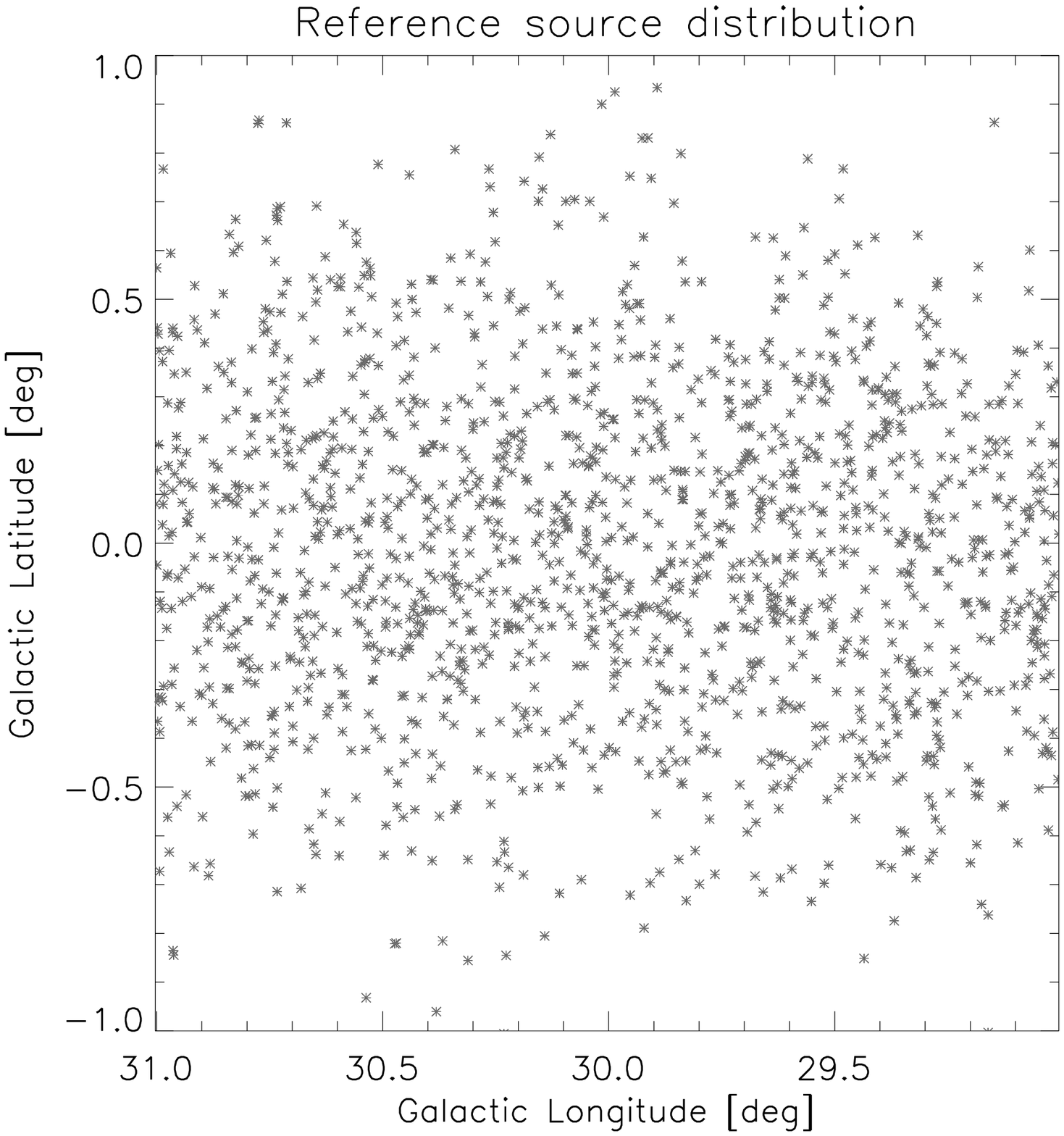}
    \end{tabular}
  \caption{Spatial distribution of all entries of the band merged catalog for the SDP $l=30$ field, and the distribution of sources randomly distributed in a plane to serve as a reference, or control, distribution.}
  \label{fig:distrib}
\end{figure*}

Throughout the remainder of our source clustering analysis,  the spatial distribution of the observed sources is systematically compared to a control distribution that exhibits no clustering properties, apart from the confinement of sources within the Galactic plane. This control distribution is used as a reference to help interpret our results. It has the same spatial coverage, and contains as many sources, as the observed fields. It is generated by drawing the Galactic longitude coordinate from a constant probability distribution, i.e. source aggregates are solely of statistical origins, and the Galactic latitude coordinate from a Gaussian distribution in order to reproduce a representative source overdensity in the Galactic plane \citep{ferriere}. Sources separated by less than 6\arcsec~are removed from the control distribution to account for the limited spatial resolution of the Herschel telescope at 70~$\mu$m. This represents about 0.5\% of the total number of sources and this has a minimal impact on the shape of the distribution. Figure~\ref{fig:distrib} presents the spatial distribution of the sources detected in the $l=30$\degr~field at 70~$\mu$m as well as the associated control distribution. 

\subsection{The Distance Matrix}
\label{subsec:dist_mat}

The computation of the  \emph{distance matrix} is the initial commonality to most approaches aiming at characterizing the spatial distribution of a given data set. Given a set of $N$ sources of known positions, the element $(i,j)$ of the distance matrix is the angular distance separating the $i^{th}$ and $j^{th}$ sources. The matrix dimension is therefore $N \times N$, and it is  symmetrical with zeros along its diagonal. It contains information from the smallest to the largest scales. The inclusion of the heliocentric distance information is discussed in section~\ref{subsec:MST_dist}.

\begin{figure}
  \centering
    \begin{tabular}{c}
      \includegraphics[width=0.45\textwidth]{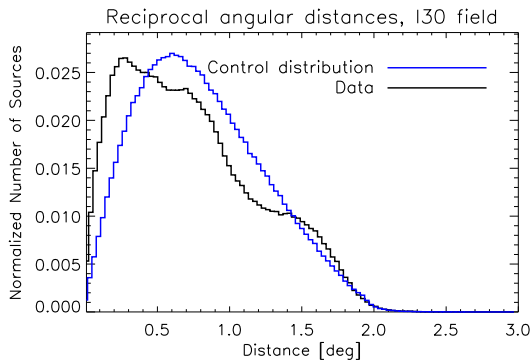}
    \end{tabular}
  \caption{Histograms of reciprocal angular distances derived from the two distributions presented in Figure~\ref{fig:distrib}. The histograms are normalized to the total number of sources in each distribution.}
  \label{fig:reciprocal}
\end{figure}

Figure~\ref{fig:reciprocal} presents the histogram of the distance matrices derived from the two distributions of Figure~\ref{fig:distrib}, and it shows that our data set exhibits an excess of short spacings compared to the control distribution, which provides the first qualitative evidence for source clustering in the observed field. The bump on the solid black line at $\theta \sim 0.75$\degr~represents the average source spacing that exists between the two star forming regions W43 and G29 which host a large fraction of the detected sources. 

\subsection{Minimum Spanning Tree and Cluster Identification}
\label{subsec:mst}

\begin{figure*}
   \centering
     \begin{tabular}{cc}
       \includegraphics[width=0.45\textwidth]{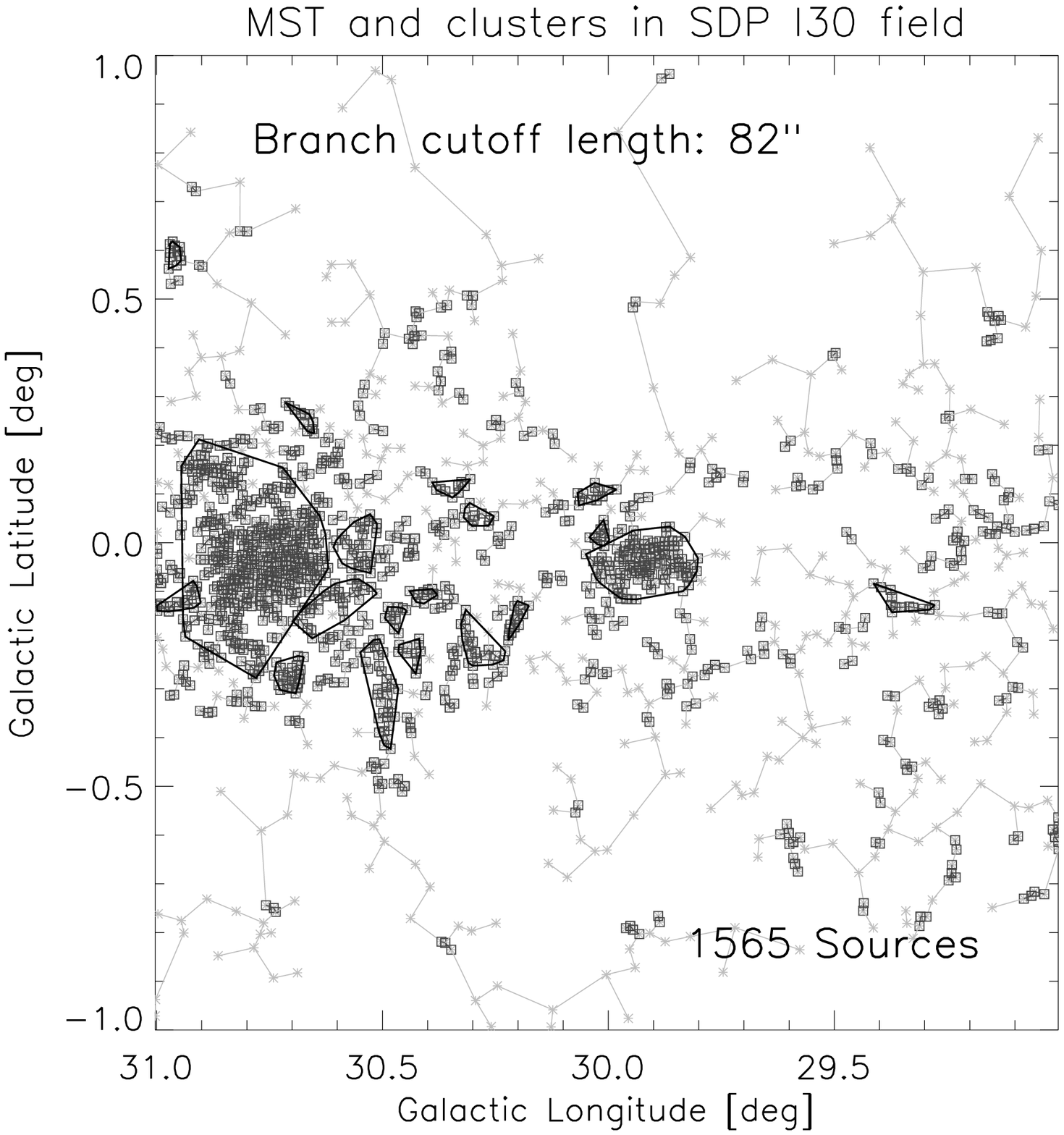} & 
        \includegraphics[width=0.45\textwidth]{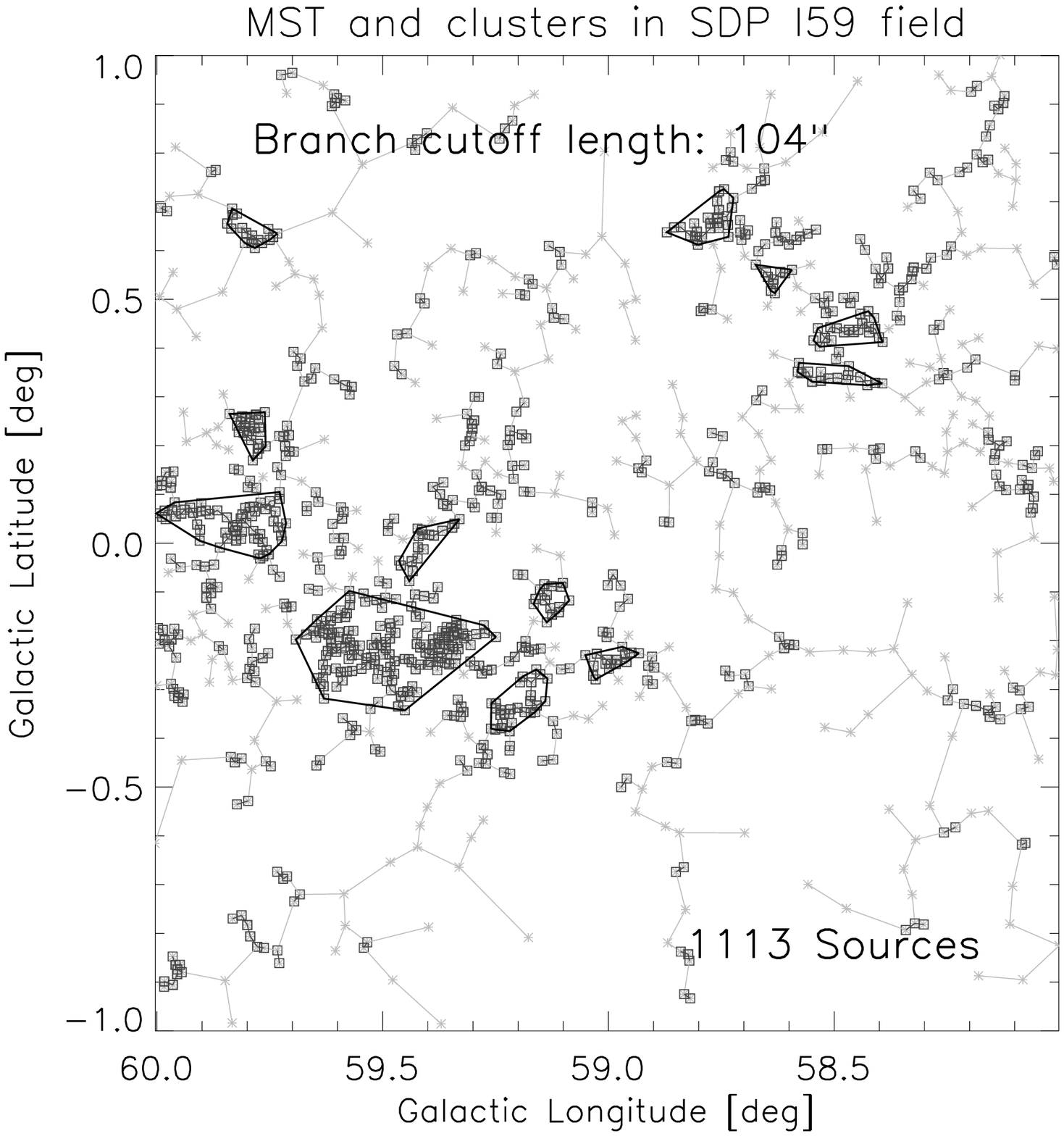}\\
       \includegraphics[width=0.45\textwidth]{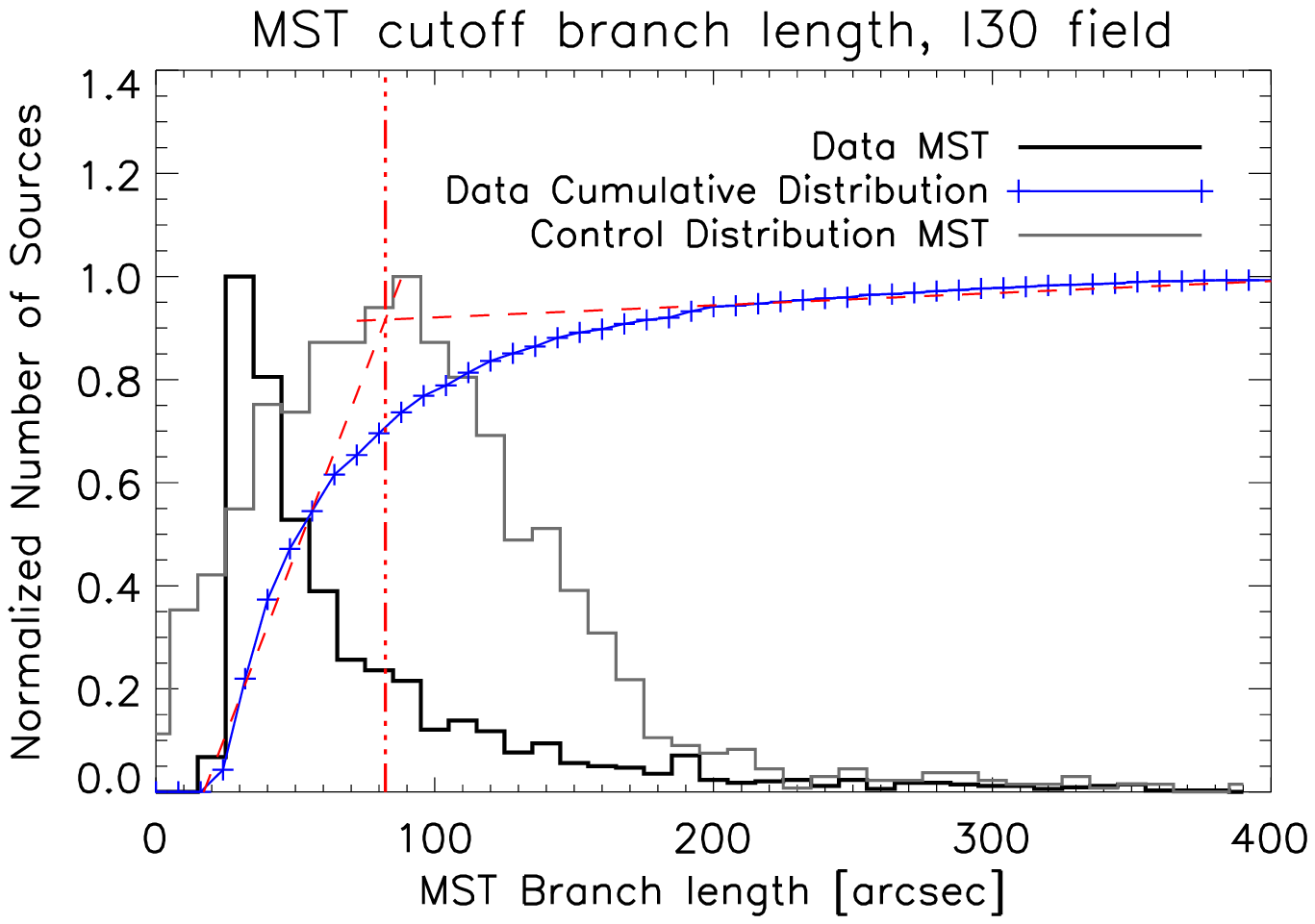} &  
       \includegraphics[width=0.45\textwidth]{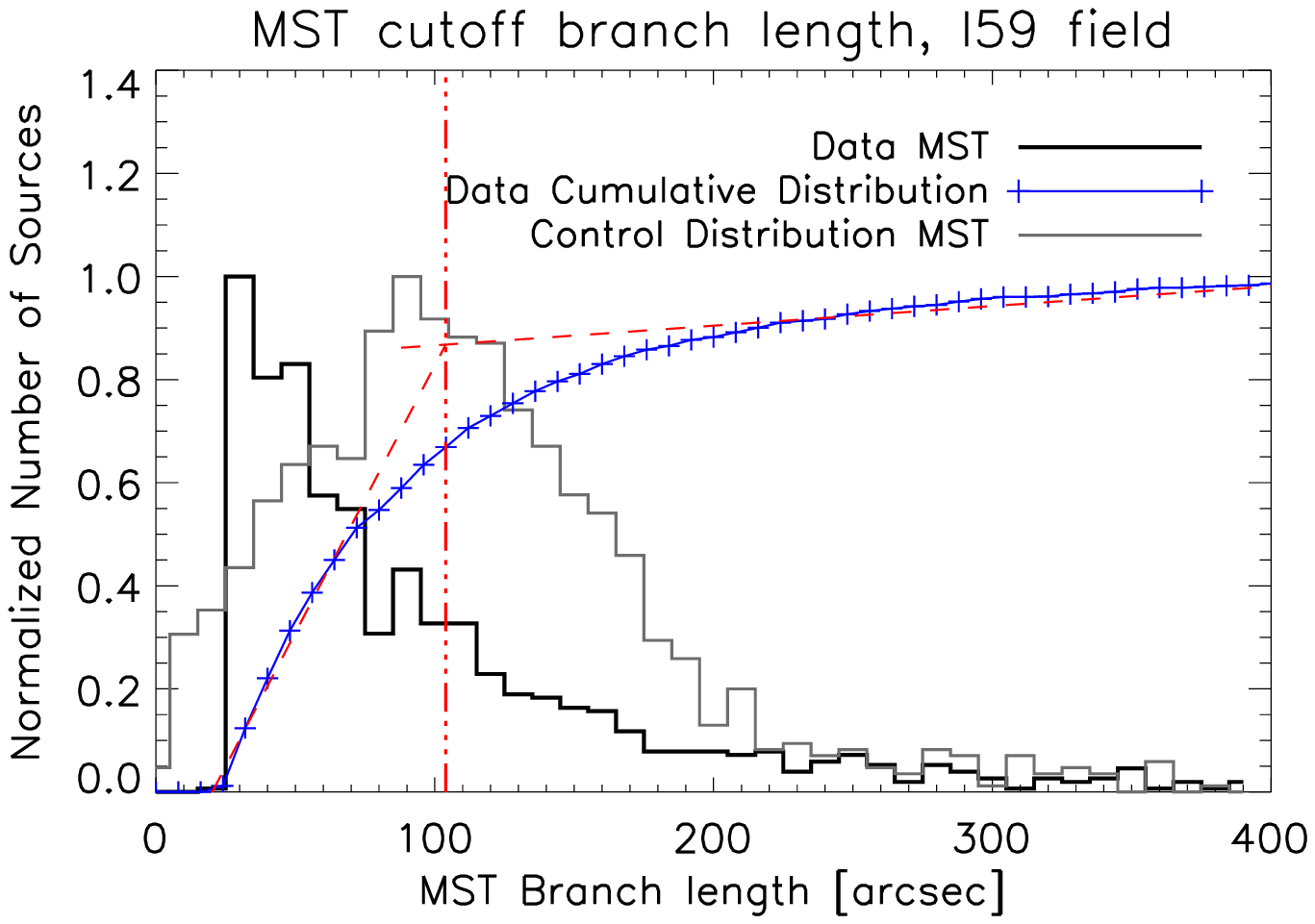}
     \end{tabular}
   \caption{Top panels: Minimum Spanning Trees derived from the band merged catalog for the  $l=30$\degr~(left) and  $l=59$\degr~(right) fields. The total number of sources considered here is given at the bottom right of the plots. The light grey segments are the branches of the MST that connect all the detected sources (light grey asterisks). Dark grey squares represent sources that are connected by branches shorter than the critical branch length. Clusters of sources containing more than 10 members are encircled by a convex hull made of black segments (see text for details). Bottom panels: Histograms of the branch lengths for the derived MSTs, as well as for the control distribution for comparison. The cumulative distribution functions are fitted by straight lines of the lower and upper parts of the branch length scale \citep[see ][]{gutermuth}, and their intersection (vertical dotted-dash line on the plot) defines the critical branch length.}
  \label{fig:mst}
\end{figure*}

A minimum spanning tree is \emph{`grown'} by connecting a set of points by a network of lines, or \emph{branches}, and by minimizing the total length of the branches while making no closed loops in the connections. There is a unique solution to this mathematical problem as long as each spacing in the distance matrix has a unique value. We use a custom IDL routine based on the Prim's algorithm to generate MSTs from the Hi-GAL source catalogs. The top panels of Figure~\ref{fig:mst} present the minimum spanning trees we have derived using all entries of the band-merged catalogs for the $l=30$\degr~and $l=59$\degr~fields, respectively. Source aggregates contrast well with the low source-density background, and we identify groupings of objects by following the definition given in \citet{gutermuth}. We exploit the distribution of MST branch lengths to define a critical length: all the sources connected by branches shorter than the critical length are assumed to belong to a group, or a source overdensity where sources are closely spaced, otherwise they are assumed to be isolated. The determination of the critical length is somewhat arbitrary since it is not based on physical assumptions, however  \citeauthor{gutermuth} used simple test case models to show that cluster-like structures can be isolated from a low-density source distribution when a critical length is derived by fitting the cumulative distribution function (CDF) of MST branch lengths with segments, and by choosing the critical length as the intersection point of the linear fits to the lower and upper ends of the branch length scale. 

The bottom panels of Figure~\ref{fig:mst} show the CDF of the MST branch lengths derived for the observed fields as well as for the control random distribution described in section~\ref{subsec:dist_mat}. The control distribution CDF appears to be more symmetrical and to peak at longer branch lengths than the observed sources, which means that short MST branches are more numerous in the observed fields, i.e. that Hi-GAL sources are more clustered than randomly distributed sources. Note also that the minimum source spacing possible in the Hi-GAL fields is set by the spatial resolution of the telescope, while it is set manually in the control distribution, so that MST branch lengths can only populate the histograms down to 6\arcsec. The observed CDF is well fitted by segments, and the critical length is found to be quite similar for both fields, $\sim$80\arcsec~and 100\arcsec~for the $l=30$\degr~and $l=59$\degr~fields, respectively. Nevertheless the shorter critical length for the $l=30$\degr~field can be attributed to the higher source density found around W43 and G29 compared to Vul~OB1 in the $l=59$\degr~field, which makes the CDF steeper at the shorter end of the branch lengths and thus displaces the intersection of the fitted segments toward shorter spacings.

\citet{bressert} give a brief summary of the various methods recently used in cluster identification, and they point out that the derived fraction of clustered sources can vary substantially for different methods (from 40 to 90\%). In our case, we follow \citet{gutermuth} and we define a \emph{cluster} as being a group of sources connected by branches shorter than the critical length, and that contains more than 10~members. We find that $\sim$70\% of the sources are associated to a cluster. \citeauthor{gutermuth} find very similar clustered fractions in nearby young star-forming clusters. The cluster identification method appears to be quite reliable for our data set as it picks up all the overdensities that can be recognized by eye, e.g. W43, G29, and Vul~OB1 for the largest associations, but also smaller substructures. The identified clusters are encircled by convex hulls (see section~\ref{subsec:cluster_charac}) in Figure~\ref{fig:mst}. 

We have also tried the alternative approach presented in \citet{battinelli}, and more recently in \citet{koenig}, to compute the critical branch length by deriving the number of clusters found in a given data set as a function of the cutoff branch length. For instance, starting from a cutoff length equal to the shortest branch of the MST, we find a single group with 2 members, i.e. the two sources that exhibit the shortest spacing, and then as the cutoff length increases and loosens the constraint to detect clusters, more clusters are found. The number of clusters is then expected to decrease when the cutoff length increases further due to the coagulation of groups into very few large clusters until the cutoff length is large enough to find a single cluster encompassing all the sources in the field. This curve therefore reaches a maximum, and \citet{battinelli} argues that the cutoff length for which the maximum is reached is the critical length to consider for the MST as it accomplishes the requirement of maximum information. \citet{koenig} find a smooth bell-shaped function for the W5 region with a well defined peak. However the curve we derive for the Hi-GAL sources has multiple local maxima rendering the definition of a critical length rather uncertain. We thus conclude that this approach is not suitable for our data set, presumably due to the presence of separate star forming complexes in the field at various distances combined with the finite spatial resolution of the observatory as pointed out by \citet{bastian07}. We therefore expect these fluctuations to get worse when considering the whole Hi-GAL survey.

\subsection{Including the distance information}
\label{subsec:MST_dist}

The distance estimate, available for over 90\% of Hi-GAL sources \citep{russeil}, is a crucial piece of information for our analysis of clustering properties. In particular, we exploit distance estimates to separate sources along the line of sight, and to convert angular distances in the sky into linear distances. 

Our initial approach was to consider all sources with a distance estimate, compute Euclidean distances between those sources in 3-D space, and then create three-dimensional minimum spanning trees. This method would in theory be the most appropriate to recover source clustering properties since it offers the best rejection of fortuitous associations due to projection effects. However, in practice, it fails to give satisfactory results because of the relatively large uncertainties associated with the estimated radial distances, $\sim$0.6-0.9~kpc (Russeil et al., private communication), compared to the tangential source spacings.

\begin{figure*}
  \centering
    \begin{tabular}{cc}
      \includegraphics[width=0.45\textwidth]{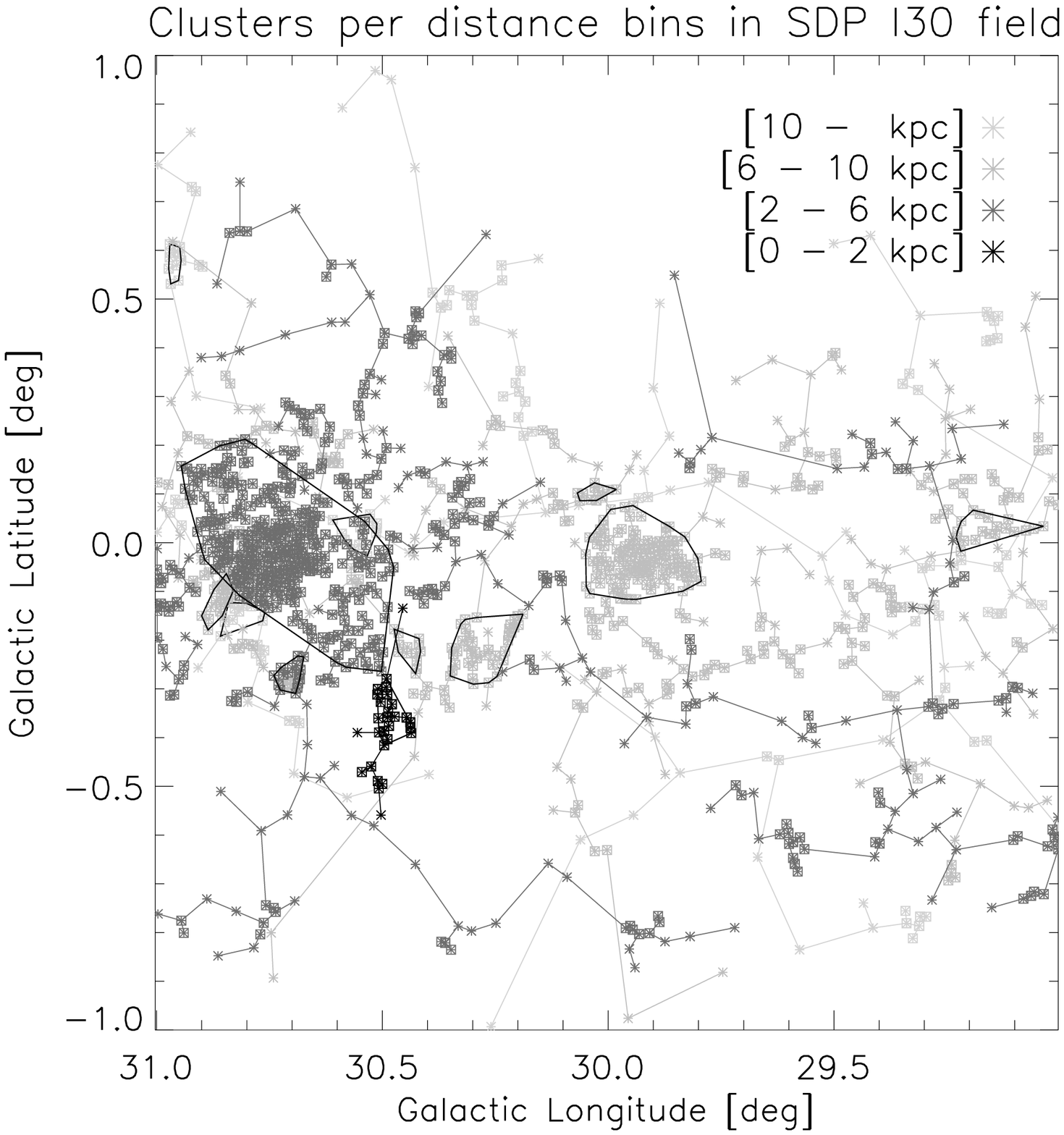}&
      \includegraphics[width=0.45\textwidth]{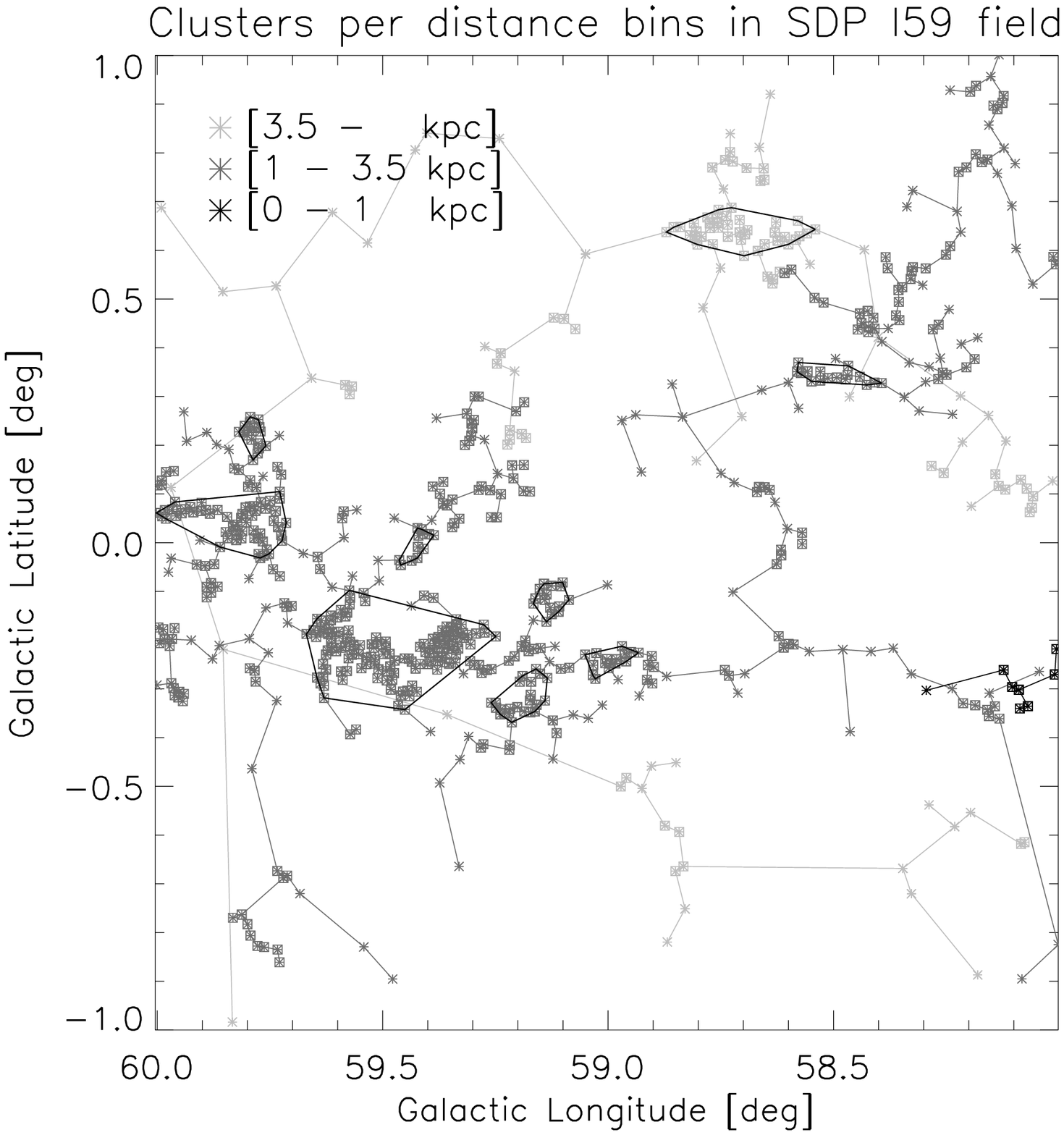}\\
      \includegraphics[width=0.45\textwidth]{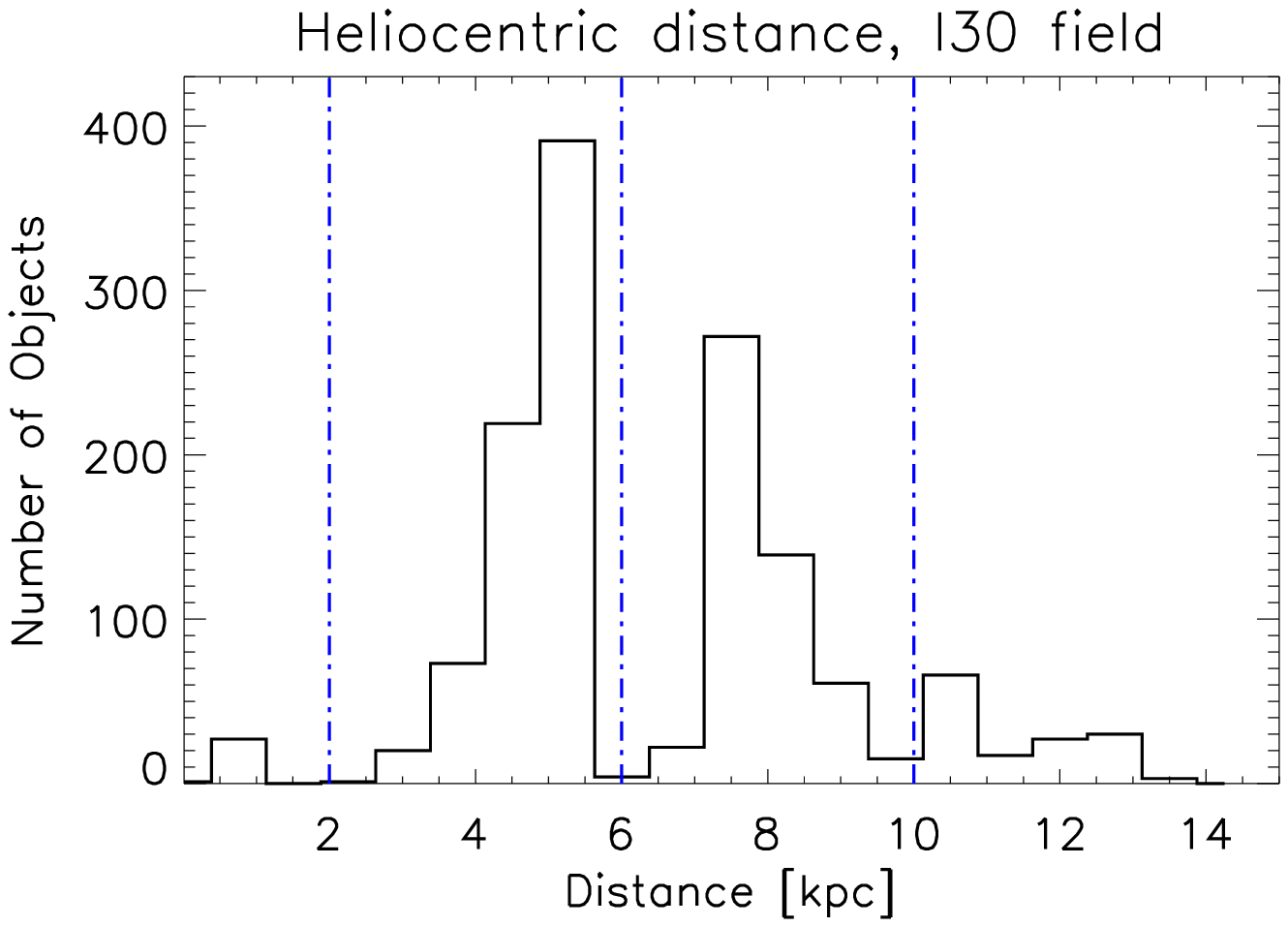}&
      \includegraphics[width=0.45\textwidth]{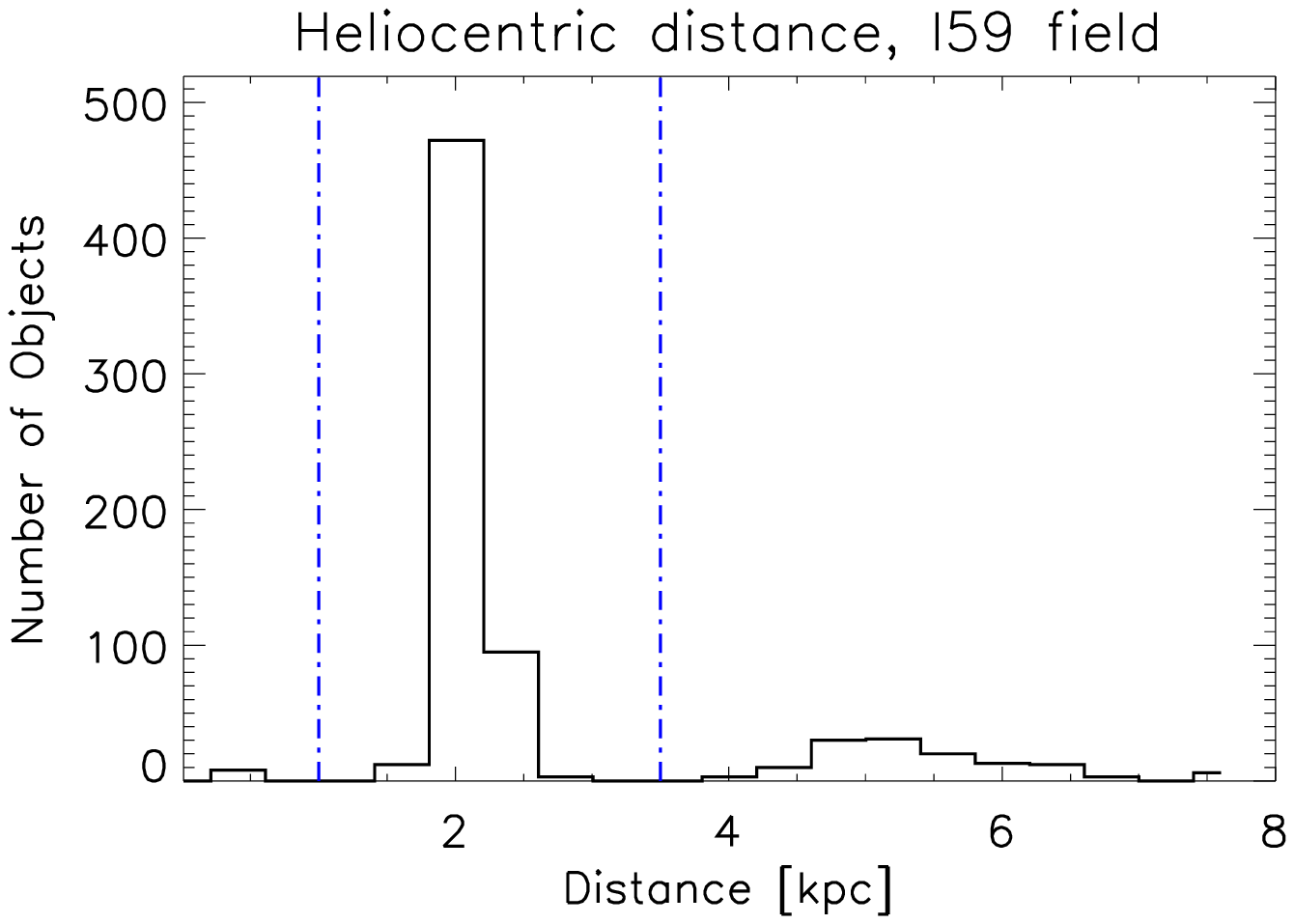}
    \end{tabular}
  \caption{Minimum Spanning Trees derived per heliocentric distance bins. Bottom panels: Histograms of source heliocentric distances for the two SDP fields. Sources belong to distinct arms of the Galaxy and can therefore be segregated per distance bins (vertical dot-dash lines define the distance bins). Top panels: Clustering properties are extracted from MSTs derived independently per distance bin (distances are grey-scale coded). The symbol convention is as in Figure~\ref{fig:mst}.}
  \label{fig:MST_perDist}
\end{figure*}

We therefore opted for a more pragmatic approach exploiting the fact that Hi-GAL sources are mainly found within the spiral arms of the Galaxy \citep{russeil} to segregate sources per heliocentric distance bins, and derive independent MSTs and cluster properties for each bin. The bottom panels of Figure~\ref{fig:MST_perDist} show the distribution of heliocentric distances in each field. The histograms possess well defined peaks that trace source overdensities located in the Galactic arms. The top panels present the 2-D MSTs and identified clusters per distance bins. The major improvement with this approach comes mainly for the $l=30$~\degr~field, for which the line of sight crosses 3~spiral arms. Indeed we find neighboring, and even overlapping, clusters that belong to different spiral arms and are actually several kiloparsecs apart. Including the distance information thus allows a clear separation of the clusters, which was not possible when treating the problem in two dimensions as in Figure~\ref{fig:mst}.

\section{Results and Discussion}
\label{sec:results}

\subsection{Cluster Characterization}
\label{subsec:cluster_charac}

Following the formalism of \citet{gutermuth}, we compute the morphological properties of clusters identified per distance bin (cf section~\ref{subsec:MST_dist}). Results are listed in Table~\ref{tab:cluster_stat} for each SDP field. 

The coordinates assigned to a cluster is the median value of the individual cluster members coordinates. Similarly, the heliocentric distance to the cluster is derived as the average distance to individual cluster members. The associated standard deviation is also given in Table~\ref{tab:cluster_stat} and represents the source spread along the line of sight, it is thus a rough indicator of the depth of the cluster.

The circular radius, $R_{circ}$, is calculated as half of the largest distance between any two members of a cluster. In other words, it is the radius of the minimum area circle that encloses the entire grouping. In addition, to account for the non-circular geometry of most clusters, we compute an effective area, $A_{hull}$, by drawing a convex hull\footnote{The convex hull is derived using the \emph{triangulate.pro} IDL routine, which computes the minimum area polygon that contains a set of points such that all internal angles between adjacent edges are less than 180\degr.} around each grouping, and by computing the associated effective radius $R_{hull}=\sqrt{A_{adjusted}/\pi}$, where $A_{adjusted}=A_{hull}/(1-n_{hull}//n_{total})$ is the adjusted effective area of the hull, $n_{hull}$ is the number of sources located on the hull, and $n_{total}$ is the total number of sources in the cluster. According to \citet{schmeja06}, the use of $A_{adjusted}$ is more appropriate than $A_{hull}$ because it has been slightly enlarged to account for all the sources located on the hull, i.e. vertices that are not strictly enclosed in the polygon. We also derive the quantity $R_{circ}^2/R_{hull}^2$ as an estimator of the cluster aspect ratio, and $A_{adjusted}/n_{total}$ as the mean surface density of source in the cluster. The last entry of Table~\ref{tab:cluster_stat} is the median MST branch length measured in the clusters, and converted from arcminutes to parsecs using the estimated cluster heliocentric distance. We discuss the relevance of this quantity to study the imprint of gravitational fragmentation in molecular clouds in section~\ref{subsec:fragment}.\\

\begin{deluxetable}{ccccccccccc}\centering
\tablecolumns{10}
\tabletypesize{\footnotesize}
\tablewidth{0pt}
\tablecaption{Cluster characteristics in Hi-GAL SDP fields. \label{tab:cluster_stat}}
\tablehead{\colhead{ID} & \colhead{Glon\tablenotemark{a}} & \colhead{Glat\tablenotemark{a}} & \colhead{Number of} & \colhead{Distance} & \multicolumn{2}{c}{R$_{circ}$ / R$_{hull}$} & \colhead{Aspect} & \multicolumn{2}{c}{Source density} & \colhead{ Median }\\
\colhead{} & [Degree] & [Degree] & \colhead{sources} & [kpc] & [Arcmin] & [pc] & \colhead{ratio} & [arcmin$^{-2}$] & [pc$^{-2}$] & Branch [pc]}
\startdata
\multicolumn{11}{c}{$l=30$\degr~field}\\[1.3ex]
\hline\\[-1.5ex]
1 & 30.549 & 0.0297 & 12 & 11.6 $\pm$ 0.19 & 3.18 / 4.19 & 10.8 / 14.1 & 0.57 & 4.59 & 52.7 & 3.72 \\ 
2 & 30.826 & -0.134 & 12  & 11.2 $\pm$ 0.07 & 3.30 / 3.03 & 10.7 / 9.89 & 1.18 & 2.41 & 25.6 & 2.78 \\ 
3 & 30.959 & 0.5884 & 12  & 12.9 $\pm$ 0.05 & 2.44 / 2.20 & 9.20 / 8.30 & 1.22 & 1.27 & 18.0 & 2.70 \\ 
4 & 29.158 & 0.0250 & 19  & 8.97 $\pm$ 0.07 & 5.83 / 3.69 & 15.2 / 9.65 & 2.48 & 2.26 & 15.4 & 3.12 \\ 
5 & 29.932 & -0.038 & 129  & 8.55 $\pm$ 0.22 & 7.90 / 6.89 & 19.6 / 17.1 & 1.31 & 1.15 & 7.16 & 1.74 \\ 
6 & 30.029 & 0.1020 & 11  & 8.13 $\pm$ 0.47 & 2.59 / 1.98 & 6.15 / 4.68 & 1.72 & 1.12 & 6.27 & 2.23 \\ 
7 & 30.291 & -0.219 & 39  & 8.04 $\pm$ 0.14 & 6.15 / 4.96 & 14.4 / 11.6 & 1.53 & 1.98 & 10.8 & 2.04 \\ 
8 & 30.426 & -0.223 & 13  & 7.96 $\pm$ 0.09 & 3.07 / 2.51 & 7.13 / 5.81 & 1.50 & 1.52 & 8.18 & 2.01 \\ 
9 & 30.855 & -0.105 & 22  & 7.65 $\pm$ 0.79 & 3.67 / 2.60 & 8.17 / 5.80 & 1.97 & 0.97 & 4.81 & 1.49 \\ 
10 & 30.733 & -0.021 & 344 & 5.65 $\pm$ 0.37 & 18.3 / 12.5 & 30.1 / 20.6 & 2.13 & 1.43 & 3.89 & 1.16 \\ 
11 & 30.692 & -0.269 & 16  & 5.52 $\pm$ 0.10 & 2.57 / 2.66 & 4.13 / 4.27 & 0.93 & 1.39 & 3.59 & 1.14 \\ 
12 & 30.489 & -0.358 & 20 & 0.89 $\pm$ 0.02 & 4.09 / 3.33 & 1.06 / 0.86 & 1.50 & 1.74 & 0.11 & 0.25 \\ [1.1ex]
\hline\\[-1.5ex]
\multicolumn{11}{c}{$l=59$\degr~field}\\[1.3ex]
\hline\\[-1.5ex]
13 & 58.734 & 0.6374 & 40  & 5.48 $\pm$ 0.16 & 10.1 / 5.54 & 16.1 / 8.83 & 3.34 & 2.41 & 6.13 & 1.76 \\ 
14 & 58.512 & 0.3404 & 16  & 2.63 $\pm$ 0.14 & 5.74 / 3.25 & 4.39 / 2.49 & 3.11 & 2.08 & 1.21 & 0.90 \\ 
15 & 58.998 & -0.239 & 16 & 2.30 $\pm$ 0.00 & 3.64 / 2.83 & 2.43 / 1.89 & 1.65 & 1.57 & 0.70 & 0.67 \\ 
16 & 59.189 & -0.332 & 23 & 2.30 $\pm$ 0.00 & 4.25 / 3.86 & 2.84 / 2.58 & 1.20 & 2.04 & 0.91 & 0.57 \\ 
17 & 59.490 & -0.215 & 132  & 2.30 $\pm$ 0.00 & 12.8 / 9.33 & 8.61 / 6.24 & 1.90 & 2.07 & 0.92 & 0.54 \\ 
18 & 59.422 & 0.0145 & 12  & 2.30 $\pm$ 0.00 & 3.09 / 2.79 & 2.06 / 1.86 & 1.22 & 2.04 & 0.91 & 0.53 \\ 
19 & 59.128 & -0.117 & 15  & 2.30 $\pm$ 0.00 & 2.87 / 3.27 & 1.92 / 2.19 & 0.77 & 2.24 & 1.00 & 0.65 \\ 
20 & 59.811 & 0.0528 & 56 & 2.30 $\pm$ 0.00 & 8.81 / 6.07 & 5.89 / 4.06 & 2.10 & 2.06 & 0.92 & 0.59 \\ 
21 & 59.787 & 0.2275 & 17  & 2.30 $\pm$ 0.00 & 2.82 / 2.30 & 1.89 / 1.54 & 1.50 & 0.98 & 0.43 & 0.52 \\ 
\enddata
\tablenotetext{a}{Cluster coordinates, expressed in Galactic longitude and latitude, represent the median coordinates of individual cluster members.}
\end{deluxetable}

\begin{figure}
   \centering
     \begin{tabular}{l}
       \includegraphics[width=0.45\textwidth]{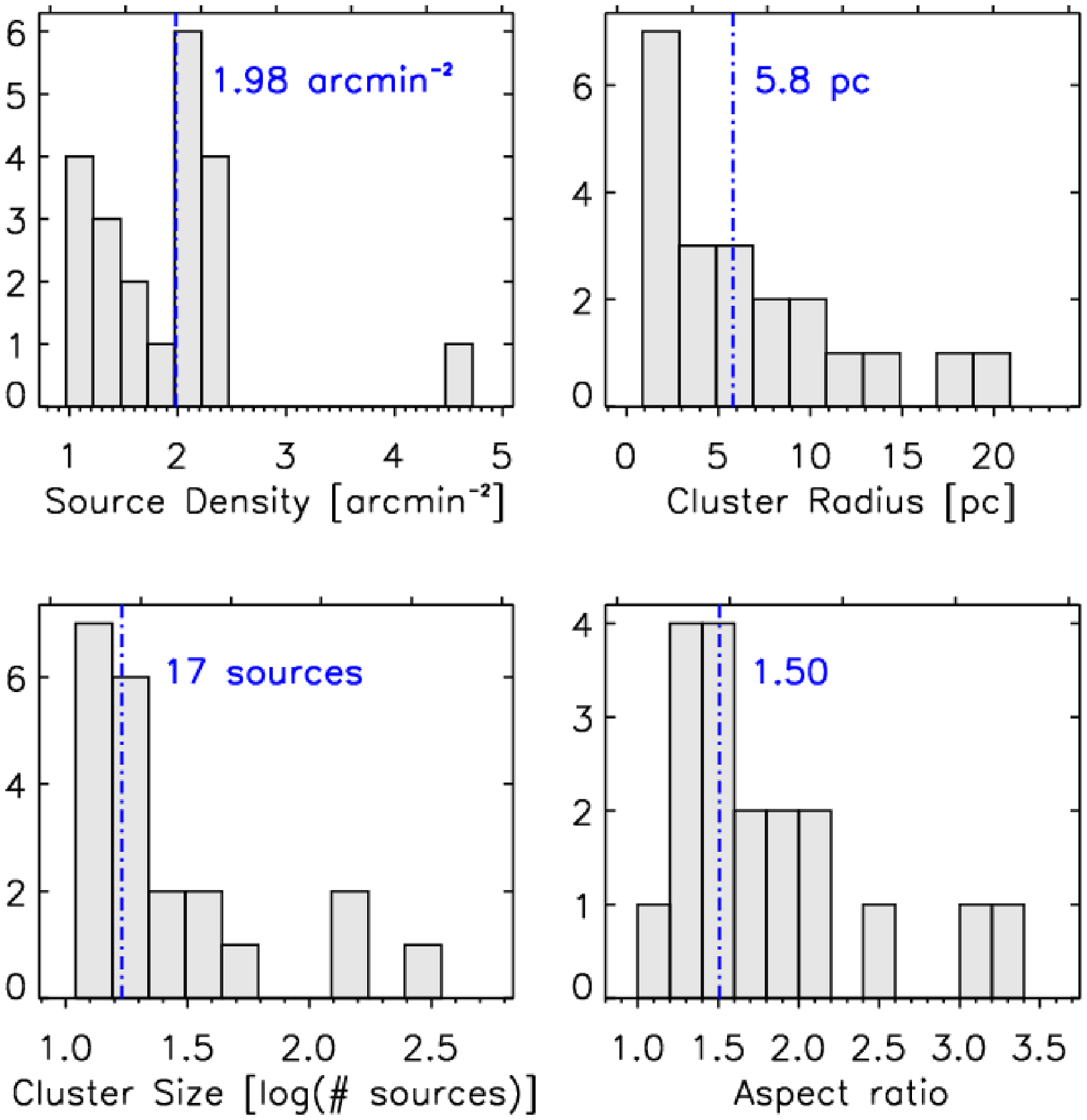}
      \end{tabular}
   \caption{Cluster core statistics derived from the MST analysis. The vertical dot-dashed lines indicate the median values of the considered quantities. Details of the computation are given in the section~\ref{subsec:cluster_charac}.}
  \label{fig:cluster_stat}
\end{figure}

Figure~\ref{fig:cluster_stat} shows the distribution of some of the quantities presented in Table~\ref{tab:cluster_stat}. We find that most clusters are small, with a median cluster size of 17~sources. The three largest groupings are located at the position of the three \textsc{Hii} regions W43, G29 and Vul~OB1. They contain over 100~sources each, which represents over half the clustered sources, and a third of all the sources in the SDP fields. The median effective radius is about 5.8~pc with a significant spread from 0.8 to 21~pc. At the low end of this range, the size is typical of long-lived gravitationally bound clusters. However large groupings, with sizes over $\sim$5-10~pc, are likely unbound structures associated with ongoing star forming activity too distant for Herschel to resolve individual substructures. The source surface density also has a significant spread in value, from $\sim$1 to 4.6~sources.arcmin$^{-2}$, with a median value of 1.98~sources.arcmin$^{-2}$. Table~\ref{tab:cluster_stat} also gives the surface density in units of sources.parsec$^{-2}$, but this physical parameter is not quite relevant for studying the most distant clusters (up to $\sim$12~kpc in our sample) due to completeness issues (we only measure the surface density of massive protostars) and spatial resolution limitations (cf section~\ref{subsec:fragment}). Most clusters tend to be elongated, with a median aspect ratio of 1.5, which property could be inherited from the primordial structure of their parental molecular cloud \citep{teixeira,allen}. In rare cases, the aspect ratio of circular small clusters can reach values below~1. \cite{gutermuth} argue that these unphysical values arise from the increasing uncertainty of $A_{adjusted}$ as the number of source members decreases and as the number of members that are convex hull vertices increases.\\

The cluster identification and characterization methods, as described above, seem to be reliable in retrieving cluster morphological parameters. Nevertheless, we consider the possible systematic effects introduced in our analysis by the choice of the 10+ cluster size limit, and the cutoff branch length. For instance, if we allow the clusters to contain fewer sources, then many more smaller clusters are detected and the parameters distribution changes accordingly. The determination of morphological parameters for large clusters is however immune to a decrease in the cluster size limit (see \citet{bastian07} for a deeper analysis). The limit of 10~members was actually chosen to be consistent with previous studies so as to make results comparable. And indeed they are: we find that 70\% of Hi-GAL sources are identified as cluster members, which matches the clustered fractions of 60-80\% found for Spitzer-identified YSOs in various Galactic star forming regions \citep[e.g. ][]{allen,koenig,gutermuth}. \\

The determination of the cutoff branch length is also somewhat arbitrary as it does not exploit any physical properties of the sources or immediate environment; kinematic measurements are the only way to establish unambiguous cluster membership. But one has to compromise to study clustering properties over very large samples ($\sim$$10^5$~sources are expected to be detected in the Hi-GAL survey once completed) and this translates into the arbitrary choice of a threshold, be it a cutoff source surface density \citep{lada,schmeja} or a MST branch length \citep{schmeja06,koenig}. \citet{gutermuth} argue that defining such a threshold from MSTs in a systematic manner for each field independently is a significant advantage over the more pragmatic definition of fixing a threshold surface density value that can potentially vary from region to region and thus miss or misidentify clusters. We therefore opted for \citet{gutermuth} approach to define the cutoff MST branch length (cf section~\ref{subsec:mst}), keeping in mind that this quantity remains somehow arbitrary.

\subsection{Cloud Fragmentation}
\label{subsec:fragment}

If we consider the idealized scenario of Jeans fragmentation in a uniform isothermal molecular cloud, gravitational instabilities can lead to the collapse and subsequent fragmentation of spherical cores when the local gas pressure can no longer support the gravitational pull of the enclosed mass. The hydrostatic equilibrium criterion can be expressed as the Jeans mass M$_J$, or equivalently as the Jeans length:
\begin{equation}
\lambda_J =\sqrt{\frac{15k_BT}{4\pi G \mu \rho}}
\label{eq:jeans}
\end{equation}
where $k_B$ is the Boltzmann constant, $G$ is the gravitational constant, $T$ and $\rho$ are the temperature and density of the cloud, and $\mu$ is the average mass per particle in the cloud. 

Assuming a typical temperature of 20~K and density of $10^5$~cm$^{-3}$ for a molecular cloud, we find $\lambda_J \sim 0.1$ pc. This quantity represents a characteristic core size which should leave its imprint on the source spatial distribution in the fragmented cloud. It is therefore to be compared with the typical source spacing found in protostellar clusters, before they migrate from their birth place and wash out their initial spatial distribution. 

\begin{figure}
   \centering
     \begin{tabular}{c}
       \includegraphics[width=0.44\textwidth]{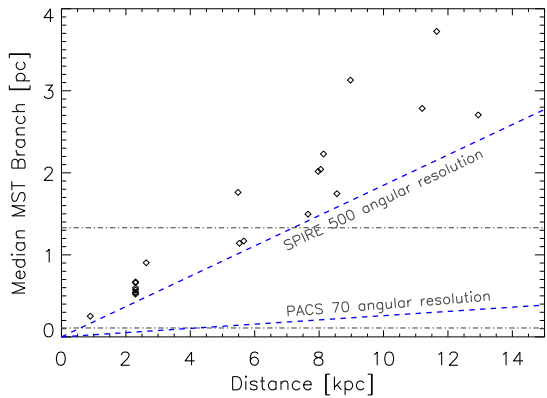}
      \end{tabular}
   \caption{Median MST branch lengths (open diamonds), derived per cluster, as a function of the cluster distance. The top and bottom dot-dash lines show the Jeans lengths for two sets of cloud physical conditions, (30~K, 10$^3$~cm$^{-3}$) and (20~K, 10$^5$~cm$^{-3}$), respectively. The two slanted dashed lines represent the angular resolving power of Herschel in the SPIRE~500 and PACS~70 bands.}
  \label{fig:cluster_stat_correl}
\end{figure}

Figure~\ref{fig:cluster_stat_correl} shows the median MST branch lengths measured in the 21 clusters identified in section~\ref{subsec:MST_dist}, as well as the Jeans lengths derived from equation~(\ref{eq:jeans}) for two sets of parameters bracketing the typical physical conditions found in the dense and cold interstellar medium, $(T,\rho) = (20~\rm{K}, 10^5~\rm{cm}^{-3})$ and $(30~\rm{K}, 10^3~\rm{cm}^{-3})$. We find that, for most clusters, the median MST branch length is larger than the Jeans lengths. To interpret this result, we compare the median MST branch lengths, as a function of the cluster heliocentric distance, with the angular resolution of the SPIRE and PACS instruments (see Figure~\ref{fig:cluster_stat_correl}), and it turns out that the typical source spacing is systematically larger than the SPIRE~500 beam size. This is an indication that the Herschel telescope cannot resolve spacings shorter than the Jeans length at kiloparsec distances in the dense interstellar medium. Completeness might also come as a limitation since we are mostly sensitive to bright and massive protostars, leaving the intermingled low-mass protostars undetected, which introduces a bias toward larger source spacings.

\subsection{Chromatic Spatial Distribution}
\label{subsec:chromatic}

From molecular condensations, to pre-stellar cores, envelop-, and then disk-accreting protostars, the peak emission from young stellar objects shifts from the millimeter to the mid-infrared regime as they evolve \citep[e.g. ][]{andre93, andre00}. The Hi-GAL survey spans over one order of magnitude in wavelengths, covering most of the above wavelength range. We therefore expect the 5-band observations to reveal young stellar objects in various stages of evolution, and possibly evidence different clustering properties for different evolutionary stages. 

\begin{figure*}
 \centering
 \begin{tabular}{c}
\includegraphics[width=0.95\textwidth]{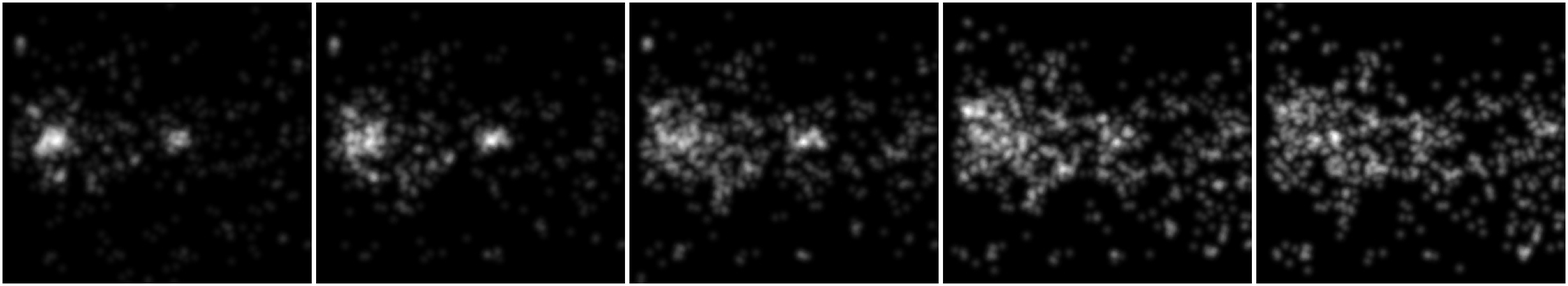}\\
\includegraphics[width=0.95\textwidth]{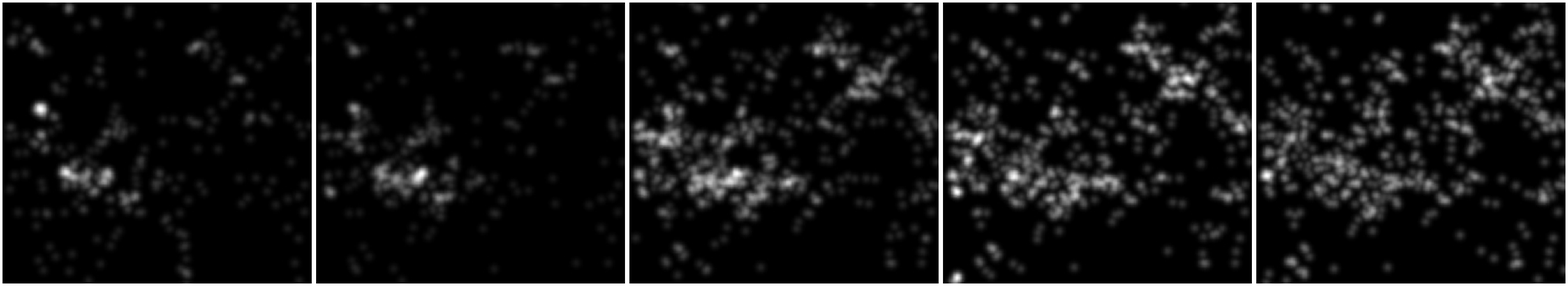}
 \end{tabular}
 \caption{Source density maps derived at all 5 Herschel bands (from left to right: detections at 70, 160, 250, 350 and 500~$\mu$m) for the 2~Hi-GAL SDP fields at $l=30^\circ$ (top row) and $l=59^\circ$ (bottom row). Short-wavelength sources are grouped in dense compact clusters while long-wavelength sources are distributed in looser and larger groups.}
            \label{fig:density}%
  \end{figure*}

We compute the source surface density from single-wavelength source catalogs to look for variations in the source spatial distribution as a function of wavelength. In practice, we use a grid of 12\arcsec-pixels, and the value assigned to each pixel is the number of sources that fall within 2\arcmin~of the pixel center. Then we convolve the resulting array with a 2D gaussian (2\arcmin~FWHM) to obtain a 2\arcmin-resolution smooth map of the source surface density.
Figure~\ref{fig:density} shows the surface density maps derived in all 5~bands for the two SDP fields. This figure illustrates the conspicuous tendency for the source spatial distribution to evolve smoothly with wavelength - in a similar way in the two SDP fields - from fairly small compact clusters at short wavelengths to looser and larger clusters at longer wavelengths. The peak surface density is 3 to 4 times higher in 70~$\mu$m clusters than in 500~$\mu$m clusters. Short-wavelength clusters seem to be preferentially located around \textsc{Hii} regions, and most clusters appear to be coincidental across the wavelengths, which is consistent with the clusters being associated with the same complex, i.e. the same parent molecular cloud. We carried out a similar MST analysis on single-wavelength source catalogs, and we found that the cutoff branch length increases monotonically with wavelength, going from 85\arcsec~at 70~$\mu$m to 180-200\arcsec~at 500~$\mu$m in the two SDP fields. The cutoff branch length determination is very sensitive to the CDF steepness at the lower end of the branch length scale (cf section~\ref{subsec:mst}), so that it can be seen as an estimator of cluster compactness. The increasing cutoff branch length with wavelength therefore confirms the evolution of clustering properties observed in Figure~\ref{fig:density}. \\

The evidence of a smooth chromatic evolution of Hi-GAL sources spatial distribution is a remarkable result, but it needs to be tested against possible observational biases that could account for the proposed trend.
We first rejected the varying angular resolution of the Herschel telescope between 70 and 500~$\mu$m as a possible cause for the observed density maps by convolving Hi-GAL maps in each band with the SPIRE~500 beam, and by detecting compact sources from these images and computing source surface density maps as in Figure~\ref{fig:density}. It appears that the smooth evolution of the source spatial distribution as a function of the wavelength remains, which confirms that the Herschel angular resolution across PACS and SPIRE bands cannot account for the observed results. 
We also considered extragalactic contamination as a potential bias in our analysis, mostly as a population of homogeneously distributed background sources at SPIRE wavelengths (250-500~$\mu$m). We ruled out this option on the ground that (1) long-wavelength source clusters are spatially correlated with the molecular gas distribution derived from $^{13}$CO integrated maps from the Galactic Ring Survey \citep{jackson}, and (2) the low extragalactic number counts from the H-ATLAS survey \citep[less than 1~galaxy per square degree is expected to be brighter than 800~mJy, ][]{clements} combined with the high confusion noise ($>$1~Jy) arising from cirrus clouds emission in the Galactic plane \citep{martin} makes it very unlikely to detect more than a few extragalactic sources in Hi-GAL fields.
Additionally, we explored the impact of the relative sensitivity in Herschel bands that might also play a role in the observed spatial distribution. We have computed source density maps similar to those in Figure~\ref{fig:density} for various subsets of single-band source catalogs by keeping only the brightest objects (flux limits were the first, second and third quartiles of the source flux distribution). It appears that the faintest sources are mainly isolated objects while the brightest sources, independently of the wavelength or field, are concentrated in the densest regions. This is consistent with the conclusions of \citet{kirk} that the most massive stars, and presumably the brightest objects, are generally located near the center of local source overdensities. However, the measured flux at a given wavelength (and heliocentric distance) is not a direct tracer of mass, but can also change with the YSO evolutionary stage. Additionally, the higher concentration of bright objects in the densest regions could partially be due to the larger number of individual objects blending into the telescope beam making the central objects appear brighter. Flux-limited source density maps are therefore difficult to interpret. \\

If the observed evolution of the spatial distribution with wavelength is genuine, then it would imply the existence of two distinct populations of objects. In particular, we argue that sources detected at SPIRE wavelength due to cold dust emission might be a mixture of stable and transient density enhancements in the ISM, which appear to exhibit moderate clustering in relatively loose associations, typically along cold filamentary structures \citep[e.g. ][]{molinari, menshchikov}, whereas more evolved objects such as protostars, which emit most of their energy in PACS bands due to the presence of accreting warm circumstellar material, are grouped in smaller and more compact clusters around \textsc{Hii} regions. We could further speculate that the feedback from massive stars on their immediate surroundings might have induced the collapse of neighboring cores, causing protostellar clusters to grow more efficiently around these \textsc{Hii} regions. These considerations are consistent with scenarios of triggered star formation. However we would need a reliable identification of the physical properties and evolutionary stages of the Hi-GAL sources, based on their SED fitting, to propose a firmer interpretation of our results. Such information should be available for later analysis of the entire Hi-GAL data set.

\subsection{Infrared Dark Clouds}
\label{subsec:irdc}

The earliest phases of star formation seem to occur preferentially in cold (T$<$20~K) and dense (n(H$_2$)$>10^{4}$~cm$^{-3}$) filamentary structures generated by gravity, turbulent mechanisms and magnetic fields  in the interstellar medium \citep[e.g. ][]{andre10,henning}. These structures are ubiquitous in the interstellar medium. In some cases, filaments are seen in absorption against the bright background in the Galactic plane, even at mid-infrared wavelengths, where the extinction is fairly low \citep{lutz, flaherty}, in which case they trace the densest structures.  Following this defining property, these objects are named Infrared Dark Clouds \citep[IRDC, e.g. ][]{carey98}. \citet{peretto} used Spitzer near- and mid-infrared surveys, GLIMPSE \citep{benjamin} and MIPSGAL \citep{carey09}, to obtain a census of IRDCs in the Galactic plane. We exploit the \citeauthor{peretto} IRDC catalog to look for associations between Hi-GAL sources and these dark filaments of cold matter. We derive the fraction of Hi-GAL sources that fall into two-dimensional ellipses generated from the morphological parameters given in the IRDC catalog (RA, DEC, major and minor axis, and position angle). We find that 32\% and 19\% of the sources in the $l=30$\degr~and $l=59$\degr~fields, respectively, are coincident with IRDCs. Among those sources, 93\% and 98\%, respectively, are classified as cluster members by the MST analysis. In contrast, only 6\% of the sources from the control distribution are coincident with IRDCs. This indicates that IRDCs are indeed associated with far-infrared sources, and thus to the earliest phases of star formation, especially with clustered sources, but it appears that a significant fraction of sources have no connections with IRDCs. The main reason for this moderate source/IRDC association rate is likely due to the fact that IRDCs do not trace systematically dense filaments in the interstellar medium, but only those that contrast well with bright infrared backgrounds. A better approach would be to rely on the sub-millimeter emission of these dense cold filaments rather than on their extinction properties. This would provide a more complete determination of dense filamentary structures, independently of the background surface brightness. \citet{menshchikov} present a qualitative examination of the associations of filamentary structures and compact objects in the Aquila and Polaris clouds, and they find that most sources lie within filaments.


\subsection{H{\small{II}} Regions}
\label{subsec:hii}

We want to quantify the observed grouping of Hi-GAL sources around \textsc{Hii} regions as mentioned in section~\ref{subsec:cluster_charac}. We use the catalog of \textsc{Hii} regions compiled by \citet{paladini} to derive the distance from each Hi-GAL source to the center of the closest \textsc{Hii} region. The $l=30$\degr~field contains 25~\textsc{Hii} regions with radii ranging from 1.6\arcmin~to 13\arcmin~and a median value of 4.3\arcmin, while the $l=59$\degr~field contains only 2~\textsc{Hii} regions. We will therefore focus our analysis on the $l=30$\degr~field due to the higher statistical significance we can reach in this field.

\begin{figure}
\centering
\begin{tabular}{c}
\includegraphics[width=0.45\textwidth]{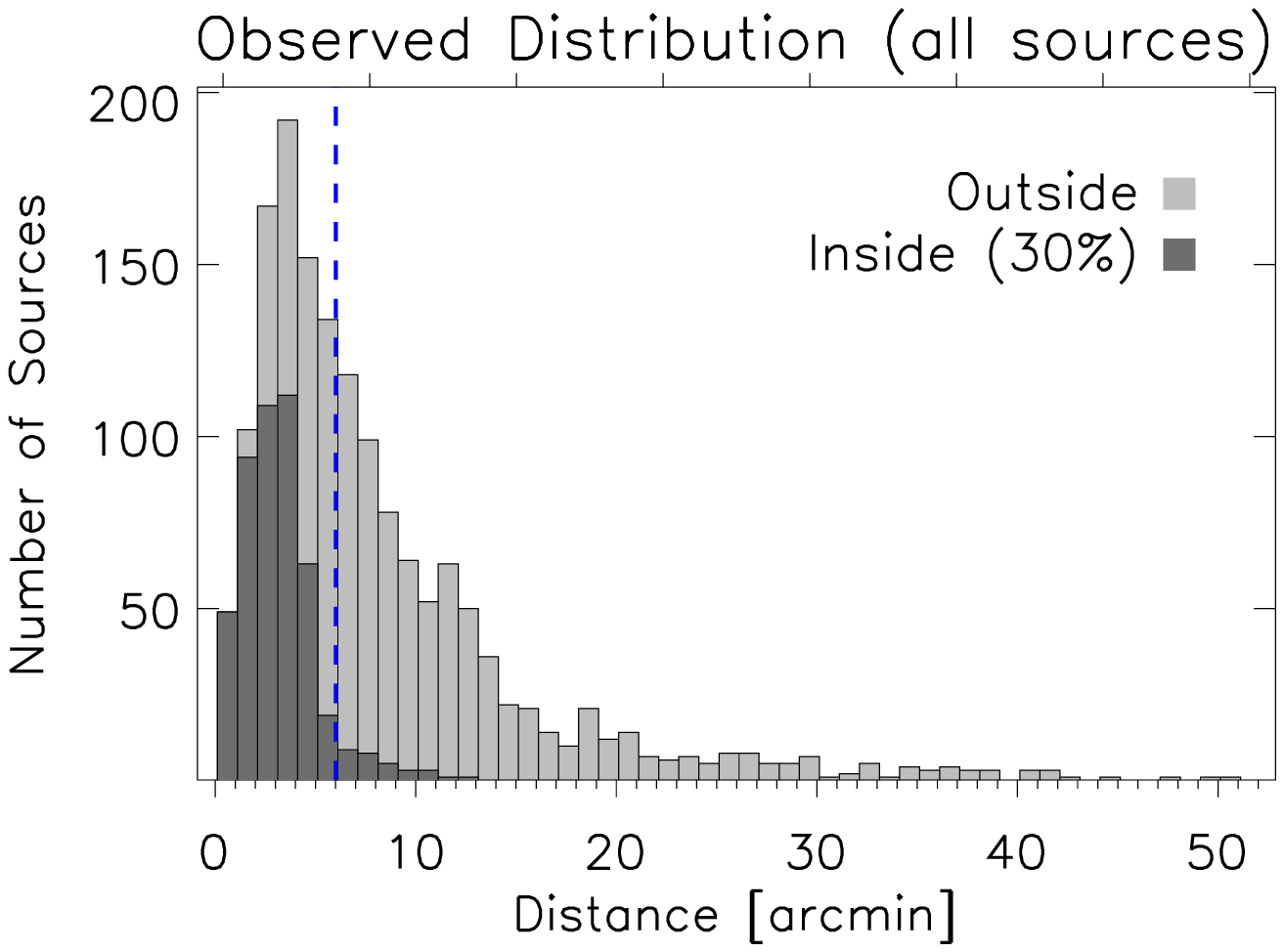}\\
\includegraphics[width=0.45\textwidth]{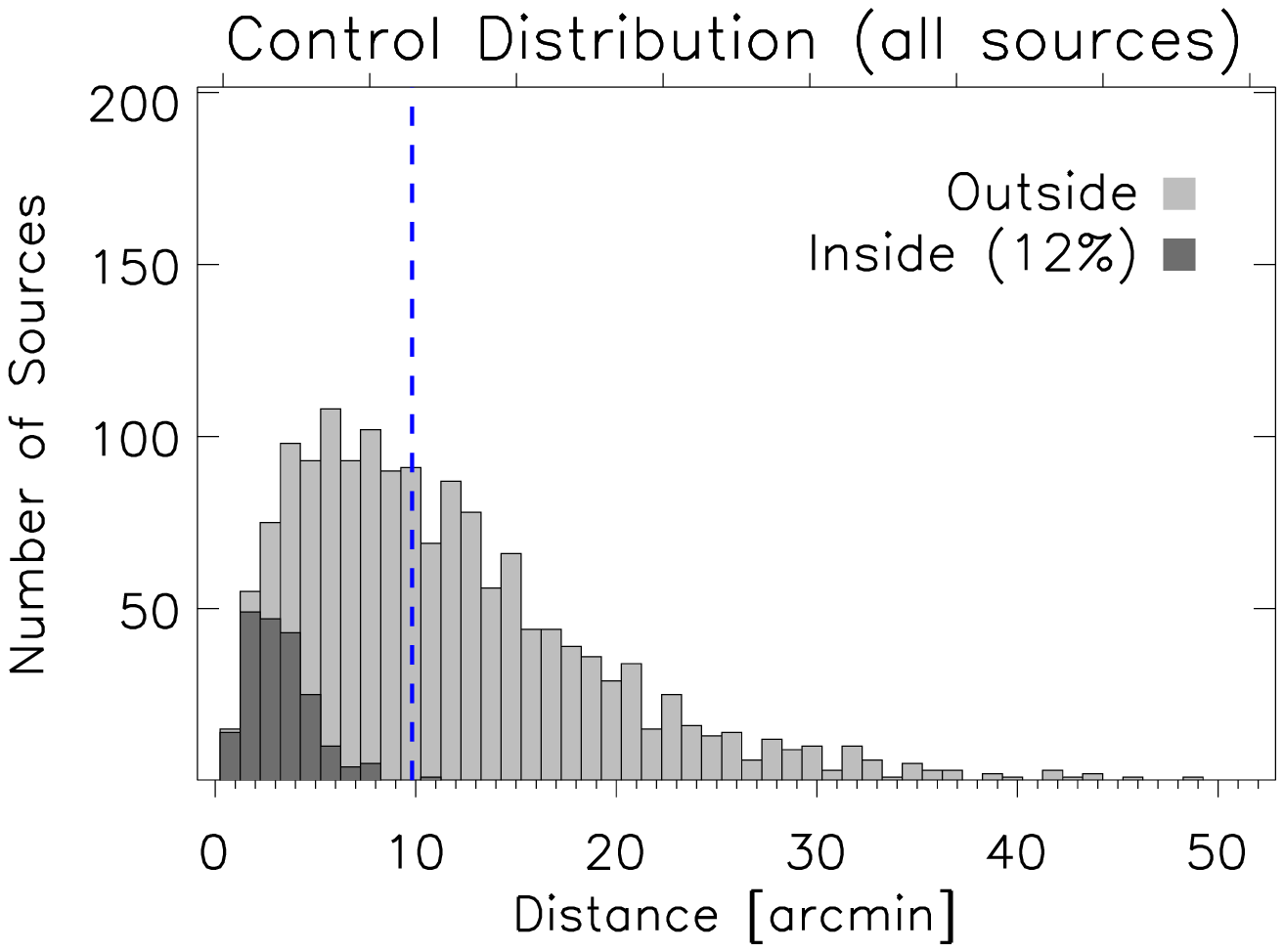}
 \end{tabular}
 \caption{Distribution of distances from HiGAL sources to the center of the closest \textsc{Hii} region in the $l=30$\degr~field, for the observed sources (top panel) and for the random control distribution (bottom panel). The vertical dashed line indicates the median distance of the distribution. The dark grey histogram indicates those sources that are located inside the \textsc{Hii} region.}
            \label{fig:hii_histo_l30}%
\end{figure}

Figure~\ref{fig:hii_histo_l30} shows the distance histograms when considering all the sources observed in the $l=30$\degr~field, as well as the random control distribution for comparison. About 30\% of the observed sources are located within a \textsc{Hii} region, and the rest of the sources appear to be distributed closer to the \textsc{Hii} regions than randomly distributed sources. In fact, over a third of the observed sources are close\footnote{The border is loosely defined as an annulus of inner and outer radii of 0.5 and 1.5 times the actual radius of the presumably associated \textsc{Hii} region.} to the \textsc{Hii} region border. Yet this occurs for only 17\% of the randomly distributed sources. Such a YSO density enhancement is consistent with the \emph{collect and collapse} scenario of triggered star formation \citep[e.g., ][]{whitworth, zavagno} in which a layer of gas and dust is compressed between the ionization and shock fronts produced by an expanding \textsc{Hii} region. Nevertheless, 55\% of the observed sources fall outside a circle 1.5 times larger than the \textsc{Hii} region radius, which means that a significant fraction of the Hi-GAL sources might not be associated with any \textsc{Hii} regions. 

\begin{figure}
 \centering
 \begin{tabular}{c}
\includegraphics[width=0.45\textwidth]{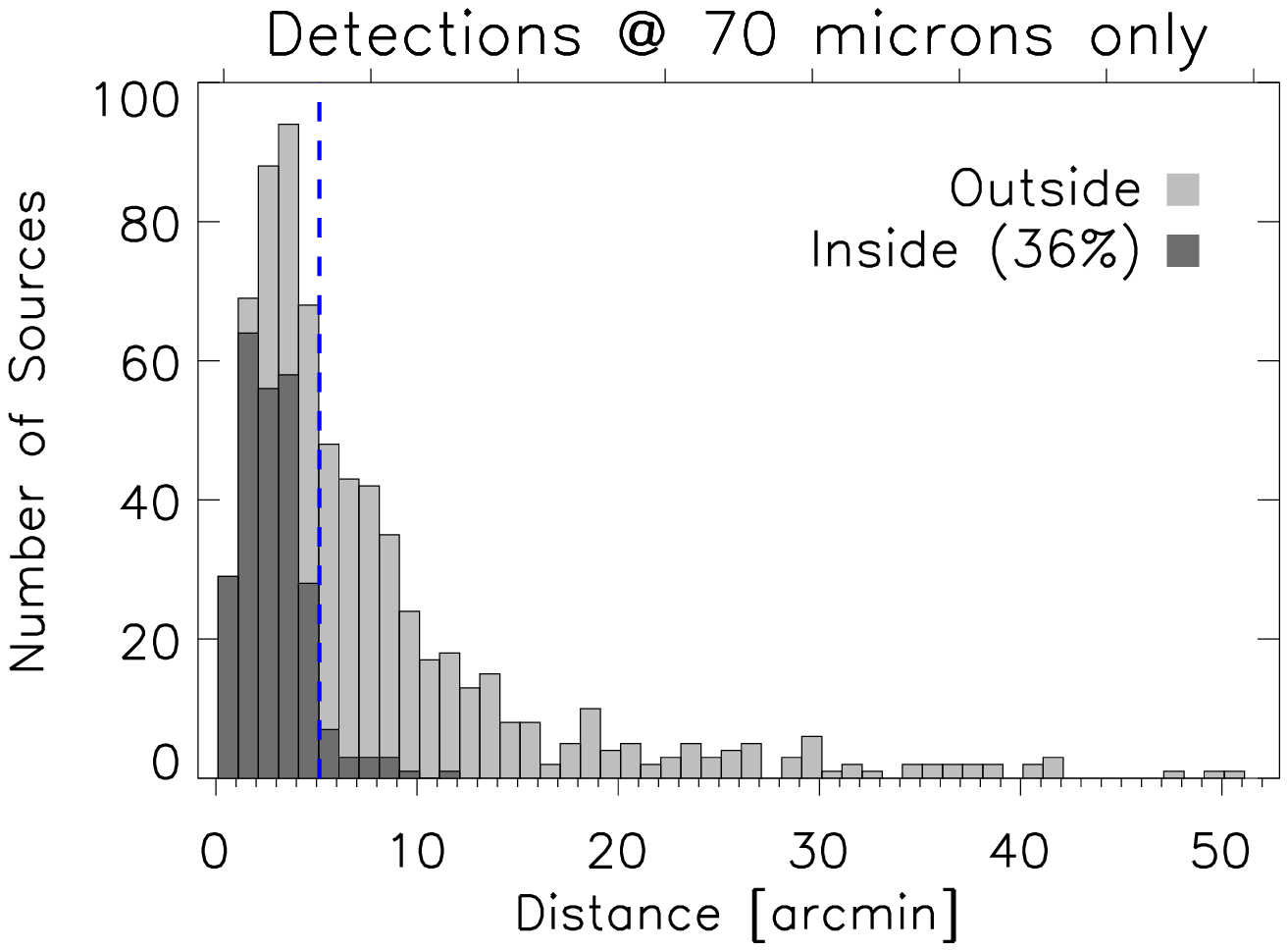}\\
\includegraphics[width=0.45\textwidth]{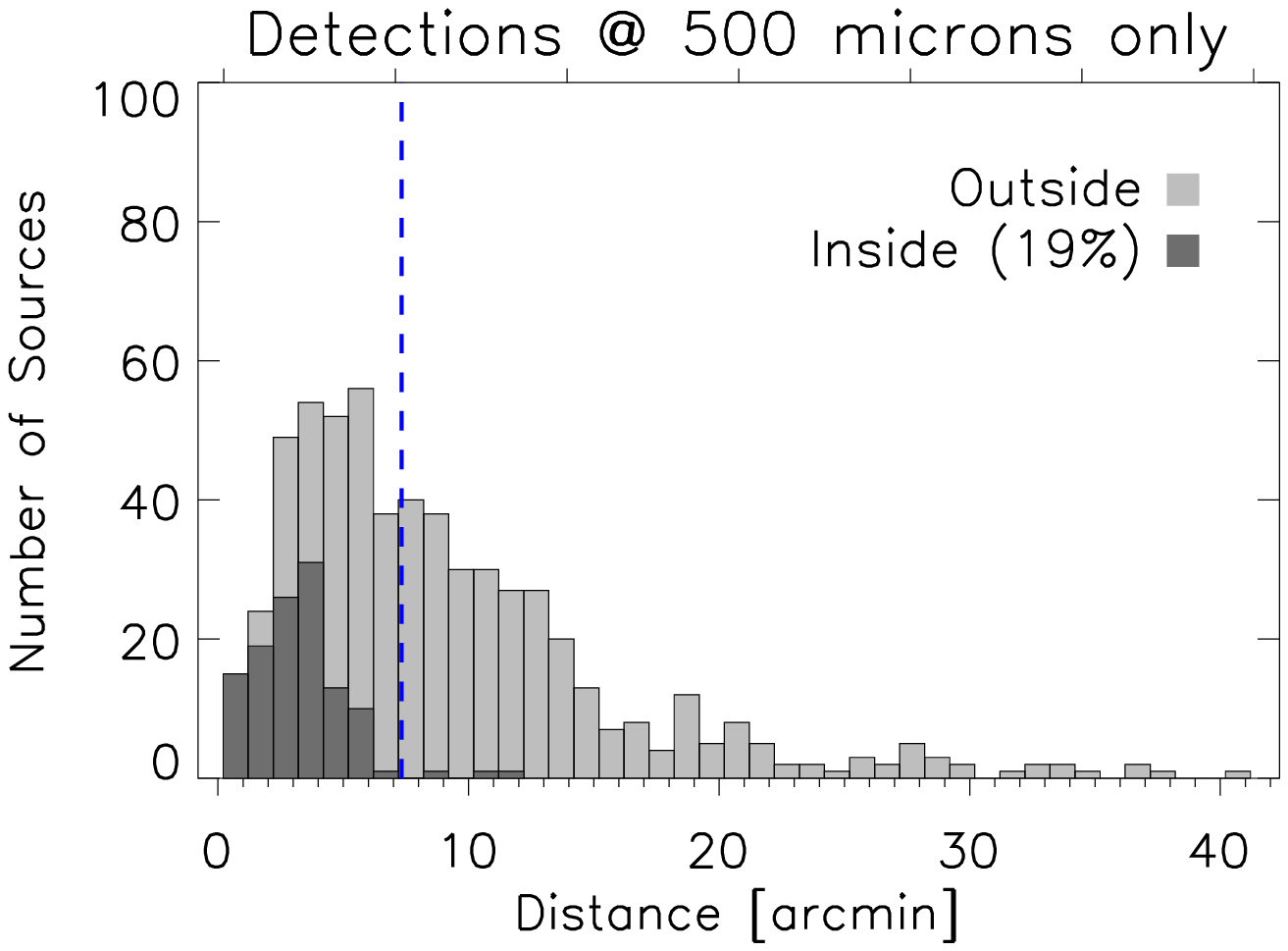}
 \end{tabular}
 \caption{Distribution of distances from HiGAL sources to the center of the closest \textsc{Hii} region in the $l=30$\degr~field, for sources detected at 70~$\mu$m (top panel) and 500~$\mu$m (bottom panel). Symbol conventions are as in Figure~\ref{fig:hii_histo_l30}.}
            \label{fig:hii_histo_l30_70_500}%
  \end{figure}

We further investigate the chromatic spatial distribution mentioned in section~\ref{subsec:chromatic} with respect to \textsc{Hii} regions. We select 70~$\mu$m and 500~$\mu$m sources from single-wavelength source catalogs, and we repeat the above analysis for the $l=30$\degr~field (cf Figure~\ref{fig:hii_histo_l30_70_500}). We find that 36\% of 70~$\mu$m sources are observed within an \textsc{Hii} region, with a median distance to the center of the region of $\sim$5\arcmin~(similar to the median radius of \textsc{Hii} regions), against only 19\% for 500~$\mu$m sources, with a median of $\sim$7\arcmin. This confirms that short-wavelength Hi-GAL sources are preferentially located within or around \textsc{Hii} regions compared to their longer wavelength counterparts.. 

\section{Summary and future work}
\label{sec:concl}

We have characterized the spatial distribution of far-infrared sources in the 2~Hi-GAL SDP fields. We first derived reciprocal distance matrices, then we built Minimum Spanning Trees including the heliocentric distance estimates necessary to separate clusters along the line of sight. Following the formalism of \citet{gutermuth} we have identified and characterized 21~clusters across the 2~fields. The 3~largest associations have more than 100~members and contain over a third of all the detected sources. The clusters are mildly elongated, with radii in the range 1 to 20~pc, and a median density of 2 sources.arcmin$^{-2}$. Half of the clusters are likely associated with \textsc{Hii} regions, and most IRDCs in the SDP fields are associated with clustered sources. However we were unable to evidence the imprint of fragmentation in molecular clouds further than $\sim$1~kpc due to the limited angular resolution and sensitivity of the survey. Nevertheless our analysis revealed the existence of 2~populations of YSOs with distinct clustering properties: short-wavelength sources tend to be clustered in dense and compact groups while long-wavelength sources are clustered in looser and larger groups, with a somewhat continuous evolution between these two clustering regimes as the wavelength increases. This remarkable result is based solely on monochromatic source density maps and minimum spanning trees, and any interpretation would be speculative at this point. We rather need to derive the physical properties of Hi-GAL sources, characterize the spatial distribution of the different classes of YSOs according to their evolutionary stage for instance, and look for similar spatial segregation effects. However the task of fitting SEDs in such crowded fields is very difficult due to the larger beam at longer wavelengths. Our team is now working on improving the reliability of SED fitting in dense clusters.

We have exploited only 3\% of the Hi-GAL survey in the present analysis. With the remaining 260~square degrees to be covered by Hi-GAL, and better tools for characterizing the physical properties of YSOs, we expect to harvest an unprecedented wealth of information with high statistical significance. We will study the cluster mass function, as opposed to the individual source mass function, and look for correlations between clustering properties and mass, evolutionary stage, environment and the galactocentric distance. In addition, the proposal for a Hi-GAL~\textsc{ii} survey has recently been accepted. It will cover $\sim270$~square degrees of the outer Galactic plane (spread by 60\degr~of Galactic longitude on either side of the Galactic anticenter). This will increase the statistics and allow us to probe a more quiescent part of the Galaxy where the star forming regime might be different than the one observed with Hi-GAL.

\acknowledgments

The authors would like to thank the entire Hi-GAL team, in particular the \emph{Distance Working Group} for their colossal effort to obtain distance estimates for the majority of Hi-GAL sources. 

{\it Facilities:} \facility{Herschel Space Observatory (PACS \& SPIRE Parallel mode)}.


\clearpage

\end{document}